\pdfoutput=1
\documentclass[a4paper,fleqn,usenatbib]{mnras}

\usepackage{newtxtext,newtxmath}
\usepackage{makecell}
\usepackage{subfigure}
\usepackage{graphicx}

\usepackage[T1]{fontenc}
\usepackage{ae,aecompl}

\usepackage{graphicx}	
\usepackage{amsmath}	
\usepackage{amssymb}	

\title[Bifurcation of equilibrium points near asteroid Bennu]{Bifurcation of equilibrium points in the potential field of asteroid 101955 Bennu}

\author[X. Wang, J. Li and S. Gong]{Xianyu Wang\thanks{E-mail:
wangxianyu12@mails.tsinghua.edu.cn (XW); lijunf@tsinghua.edu.cn (JL); gongsp@tsinghua.edu.cn (SG)}, Junfeng Li\footnotemark[1] and Shengping Gong\footnotemark[1]\\
School of Aerospace Engineering, Tsinghua University, Beijing, 10086, China}

\date{Accepted 2015 October 19.. Received 2015 September 24; in original form 2015 May 20}

\pubyear{2016}

\begin{document}
\volume{455}
\pagerange{3724--3734}
\maketitle

\begin{abstract}
The stability and topological structure of equilibrium points in the potential field of the asteroid 101955 Bennu have been investigated with a variable density and rotation period. A dimensionless quantity is introduced for the nondimensionalization of the equations of motion, and this quantity can indicate the effect of both the rotation period and bulk density of the asteroid. Using the polyhedral model of the asteroid Bennu, the number and position of the equilibrium points are calculated and illustrated by a contour plot of the gravitational effective potential field. The topological structure and the stability of the equilibrium points are also investigated using the linearized method. The results show that there are nine equilibrium points in the potential field of the asteroid Bennu, eight in the exterior of the body and one in the interior of the body. Moreover, bifurcation will occur with a variation of the density and rotation period. Different equilibrium points will encounter each other and mix together. Thus, the number of equilibrium points will change. The stability and topological structure of the equilibrium points will also change because of the variation of the density and rotation period of the asteroid. When considering the error of the density of Bennu, the range of the dimensionless quantity covers the critical values that will lead to bifurcation. This means that the stability of the equilibrium points is uncertain, making the dynamical environment of Bennu much more complicated. These bifurcations can help better understand the dynamic environment of an irregular-shaped asteroid.
\end{abstract}




\section{Introduction}

Asteroids and comets are thought to be critical to understanding the origin and evolution of the solar system, and many countries and space agencies have launched missions to these small bodies, such as the NEAR Shoemaker mission, the Dawn mission and the Hayabusa mission. Shoemaker is operated by NASA and was sent to orbit asteroid 433 Eros, with a flyby of asteroid 253 Mathide during its journey to Eros \citep{b22}. Shoemaker landed on the surface of Eros after several orbital manoeuvres, thus becoming the first spacecraft to soft land on an asteroid \citep{b27}. Dawn is also operated by NASA and has visited two asteroids, 4 Vesta and 1 Ceres, during one journey \citep{b16}. Hayabusa is operated by JAXA as an asteroid sample return mission. It encountered asteroid 25143 Itokawa in 2005 and landed on the asteroid to collect samples, which were sent back to earth after 5 years \citep{b7}. Recently, Rosetta arrived at comet 67P/Churyumov-Gerasimenko and released a lander named Philae to its surface \citep{b6}. It is the first time that a manmade machine landed on a comet nucleus, which shows great advancement for human deep space exploration and may obtain fruitful results about comets.

Though there are some small body missions being carried out, the dynamical environment near an asteroid or a comet is still full of challenge and unknown. In the Hayabusa mission, the spacecraft carried a detachable minilander, MINERVA, but it failed to reach the surface of Itokawa because of the complicated dynamical environment \citep{b26}. Most of the small bodies are irregular-shaped along with a rapid rotation period, which makes the orbital dynamics in the vicinity of them highly nonlinear and quite different from that of planets. The gravitational field of asteroid provides physical examples in the universe for nonlinear dynamical system theory, and research findings about this can help design deep space exploration tasks for small bodies. The gravitational field in the vicinity of an asteroid is very complicated and has very rich dynamical phenomena such as eject orbits and capture orbits \citep{b23}. This complexity is very attractive, and many studies have been conducted to investigate the orbital dynamics in the vicinity of small bodies.

One method to establish the gravitational field is the potential harmonic expansion method, which is widely used to theoretically analyse the orbital dynamics near an asteroid. The dynamical environment of the asteroid 4769 Castalia, such as the equilibrium points, zero-velocity curves, stable and unstable manifolds, and period orbits, was investigated using the harmonic expansion method \citep{b18}. The more precise dynamical model includes solar tide and solar radiation pressure perturbations, and the orbital motion was investigated to determine the stable orbits \citep{b17}. There are also many researchers studying the gravitational field of small bodies using equivalent models, such as a massive straight segment, a homogeneous cube and a triaxial ellipsoid \citep{b14,b15,b10,b11}. A massive straight segment was discovered to have four equilibrium points and families of period orbits, and the bifurcation was found, which shows that there are internal relations between the period orbits based on the different equilibrium points \citep{b15}. Similar results are found under the dynamical model of a triaxial ellipsoid, which is a closer model to real asteroids or comets \citep{b14}. The orbital dynamics in the gravitational field of a massive homogeneous cube were investigated by \citet{b10,b11} both with and without rotation. \citet{b25} introduced the polyhedral method to simulate the gravitational field of an irregular-shaped small body, which allows researchers to study orbital dynamics in a more precise way. The equilibrium points and their stability can be studied with more details using the polyhedral shape models of the small bodies \citep{b24}. Moreover, 29 basic families of period orbits around the asteroid 216 Kleopatra were calculated using a hierarchical grid searching method \citep{b28}.

Previous research on the gravitational field of asteroids or comets is based on certain physical properties, such as density, size and rotation period, most of which come from ground-based observations. These physical properties are not very accurate, and different references show differences \citep{b1,b2}. \citet{b8} used these differences to study the structural constrains of 216 Kleopatra with the size as a variable parameter. On the other hand, because of the Yarkovsky-O'Keefe-Radzievskii-Paddack (YORP) effect, the rotation period of some small bodies has become shorter, for example, the asteroid 54509 YORP (2000 PH5). In 2005, the asteroid rotated 250 degrees more than was expected in 2001, and its spin velocity is still rising every day \citep{b20}. Researching certain physical properties of the asteroid is meaningful for the discovery of the dynamical system, such as the equilibrium points and period orbits. Considering the uncertainty of these properties could lead to more fruitful results, which can help to understand the dynamical environments near an asteroid better. In this paper, the number and stability of equilibrium points in a gravitational field are investigated using variable physical properties. Moreover, the bifurcation of the equilibrium points is studied during the variation. The physical properties can be denoted by a dimensionless quantity that relates the bulk density and rotation rate to normalized equations of motion. The variation of this dimensionless quantity leads to fruitful results of the orbital dynamics in the vicinity of an asteroid, which indicates bifurcation for not only the number of the equilibrium points but also the topological structure of the manifold near the equilibrium points. The results of this research can help better understand the dynamical environment near a small body by considering the inaccuracy of the physical properties. The results show that the equilibrium points of the asteroid Bennu can be either linear stable or unstable because of the uncertainty of the physical properties, making the dynamical system more complicated. The study can also help design the orbits around an asteroid or a comet if a stationary orbit or a halo orbit is needed. Different physical properties can lead to different designations of the orbits for not only the position of the equilibrium points but also their stability, which changes with the variation of the physical properties. Moreover, in order to gain a global view of the dynamical system of the small bodies, both the internal and external potential fields are investigated.

In Section 2, the physical properties, which are obtained from ground-based observations, are shown as a basic reference of the variation. The equations of motion in the body-fixed frame of a small body are derived, normalized and applied to the asteroid 101955 Bennu in Section 3. The equilibrium points and their stabilities have also been studied using the linearized method around the equilibrium points. In Section 4, we investigate the stability and topological structure of the equilibrium points in the potential field of irregular-shaped small bodies with variable density and rotation period, and this analysis yields fruitful results of the dynamical system of a non-central rotating potential field. It is found that the number, stability and topological of the equilibrium points all change significantly with a variation of the physical properties. The number of equilibrium points decreases when applying an increasing dimensionless quantity. Some equilibrium points change their stability from unstable to linear stable, and thus, the topological structure also changes.

\section{GEOMETRIC PROPERTIES AND MODEL OF BENNU}

The asteroid 101955 Bennu was first observed in September 1999 and was provisionally named 1999 RQ36. It is an Apollo type asteroid with a semi-major axis of 1.126 au, its orbit crosses that of the Earth, and it belongs to the near-Earth Asteroids. Therefore, it is a potential Earth impactor and is listed on the Sentry Risk Table. According to the Sentry Risk Table, the asteroid Bennu has 78 potential impacts during the years 2175-2199. The sum of the impact probabilities from all detected potential impacts is $3.7 \times 10^{-4}$, and the highest individual impact probability is $1.1 \times 10^{-5}$ in 2180 \citep{b3}. Because of its high risk of impact to Earth, many optical observations have been obtained for Bennu from the Arecibo Observatory and the Goldstone Tracking Station, which help to obtain precise data for this near-Earth asteroid, both for the orbit and physical properties. Based on the visible spectrum, Bennu is classified as a B-class asteroid. Most of the asteroids belonging to the B-class are thought to be from the middle or outer main belt, and this categorization can help predict the origin of Bennu. These asteroids are thought to be primitive and volatile-rich remnants from the early solar system, just like the CI or CM meteorites \citep{b4}. These asteroids are also thought to be influenced by the YORP effect significantly. The rotation speed will spin up, and the rotation axis will also change. Additional observations are still needed either from the ground or from a spacecraft visiting the asteroid. Moreover, Bennu is the target of OSIRIS-REx sample return mission, which implies a further investigation of the asteroid. The orbit brings Bennu close to Earth every six years, making a series of low-cost transfers to rendezvous and sample return. It is planned that the spacecraft will arrive at Bennu in August 2018 and will acquire sample in late 2019.

The three-dimensional shape model of the asteroid Bennu adopted in this paper was driven from the radar images and optical lightcurves by \citet{b12}. The asteroid is quite irregular-shaped, has no symmetry and has a ¡°spinning top¡± shape. Fig. 1 shows the polyhedral model with 1348 vertices and 2692 faces, which illustrates the geometric dimensions of Bennu to be $576 \times 539 \times 526\ \rmn{m}$ with a mean diameter of $492 \pm 20\ \rmn{m}$. The volume is calculated as $0.0623 \pm 0.006\ \rmn{km}^{3}$ using this model.

\begin{figure}
 \includegraphics[width=41.5mm]{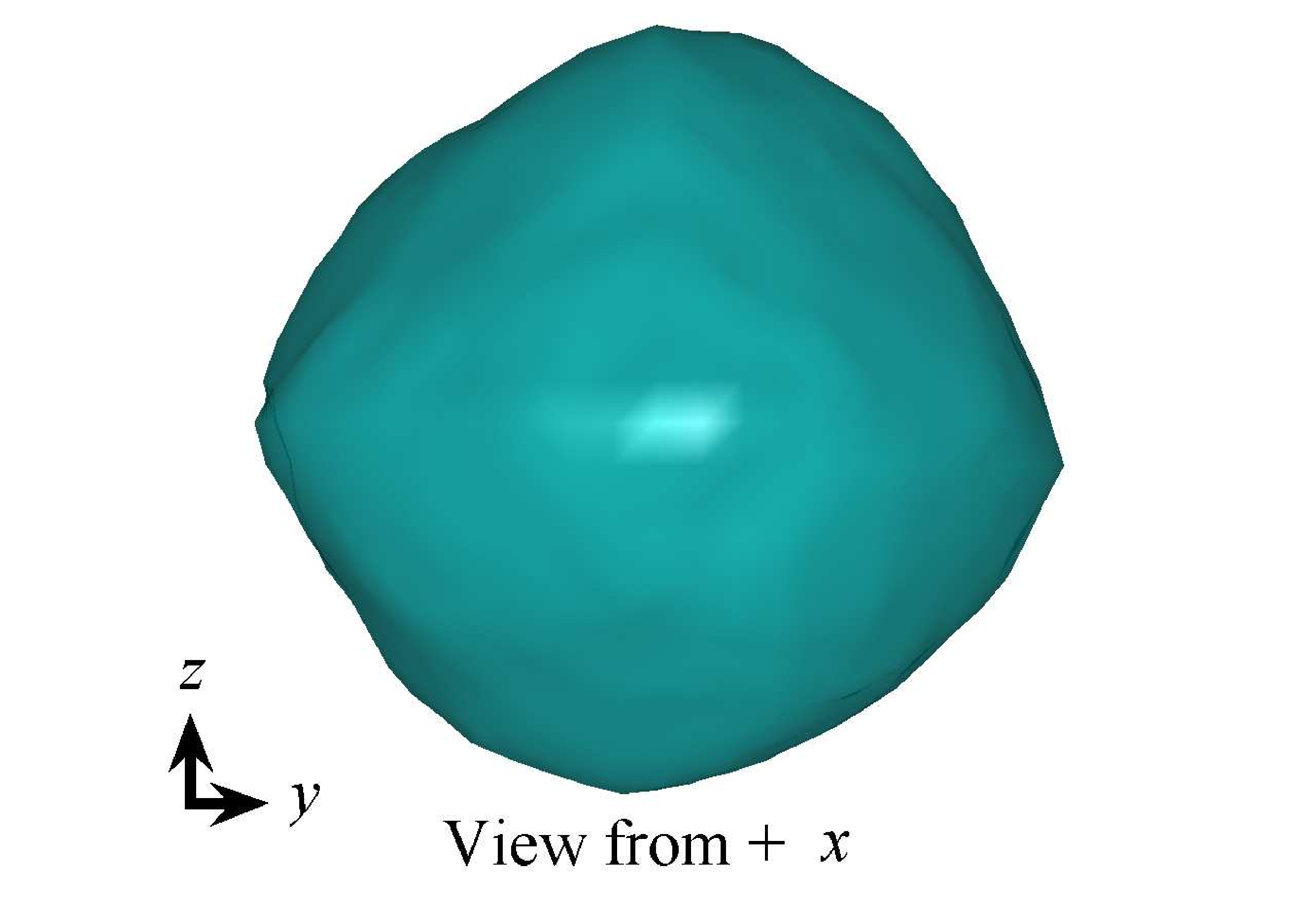}
 \includegraphics[width=41.5mm]{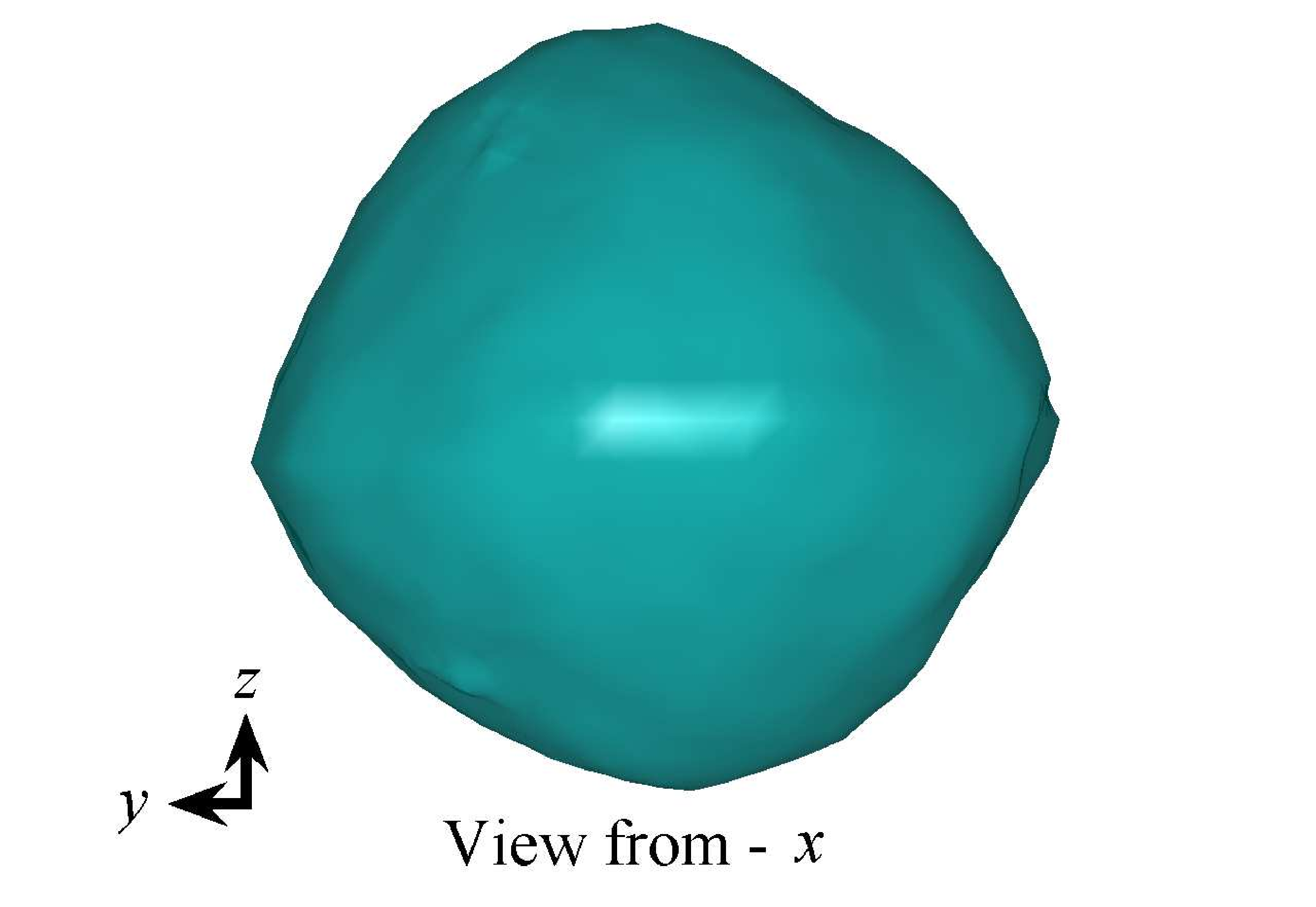}
 \includegraphics[width=41.5mm]{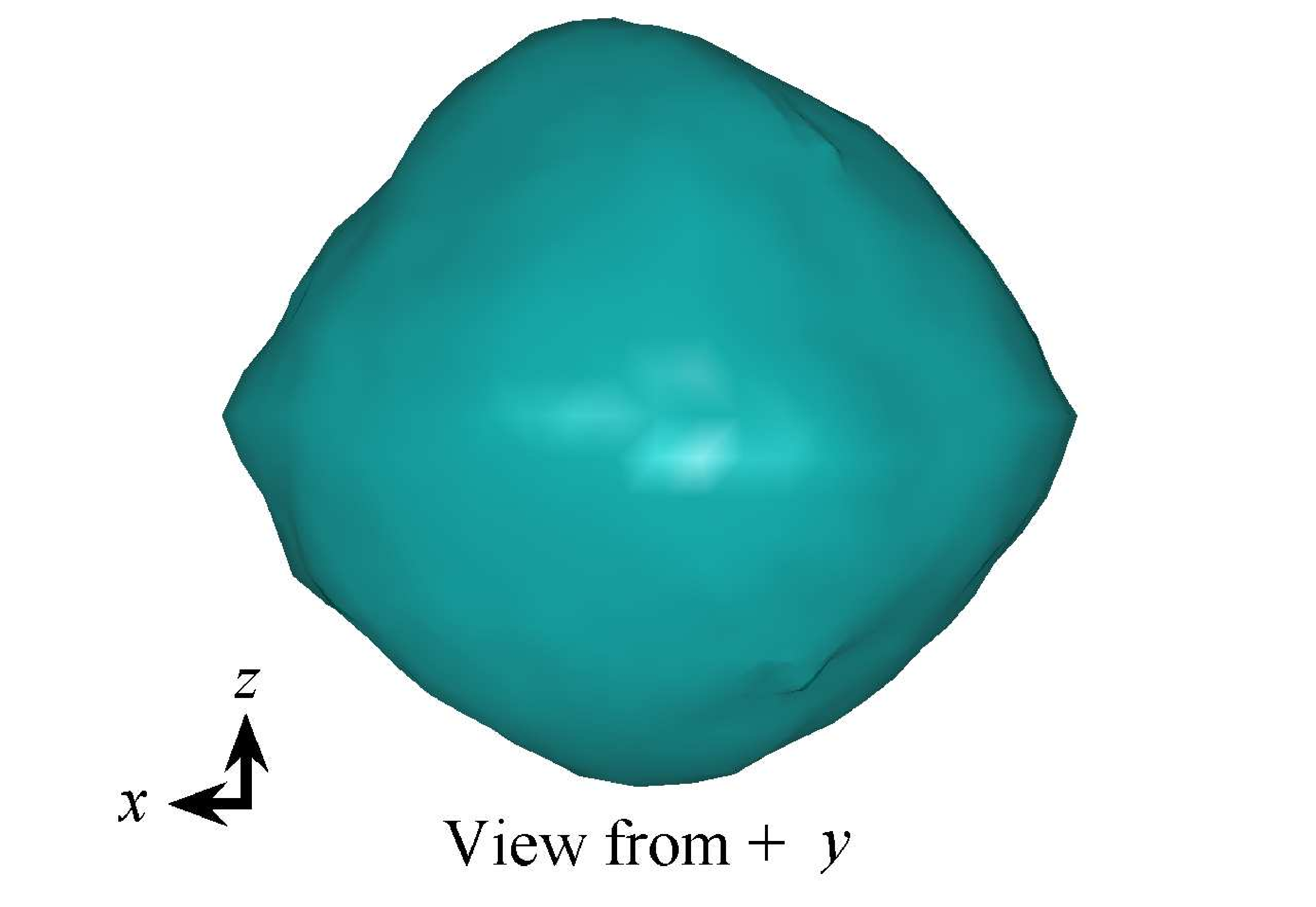}
 \includegraphics[width=41.5mm]{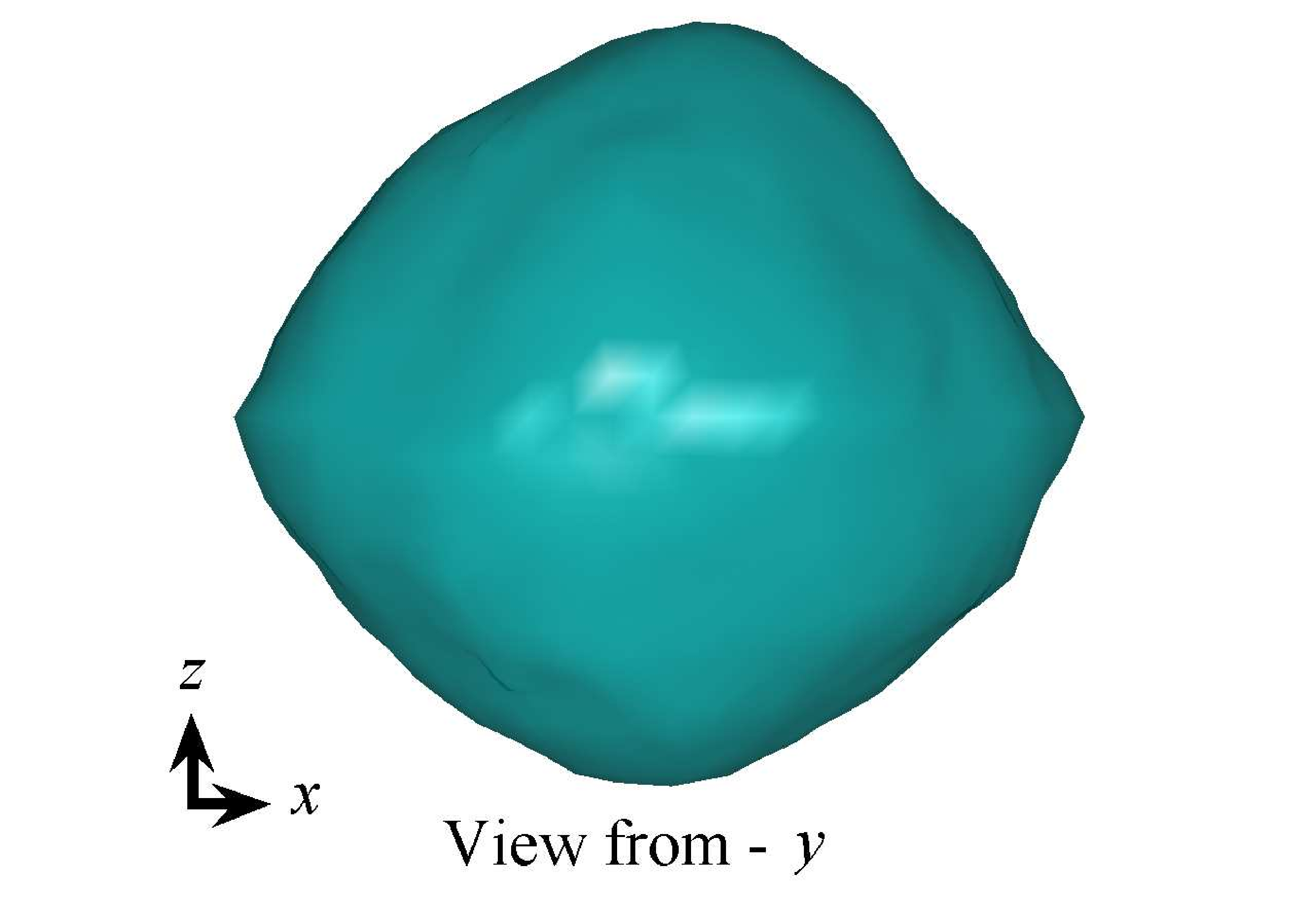}
 \includegraphics[width=41.5mm]{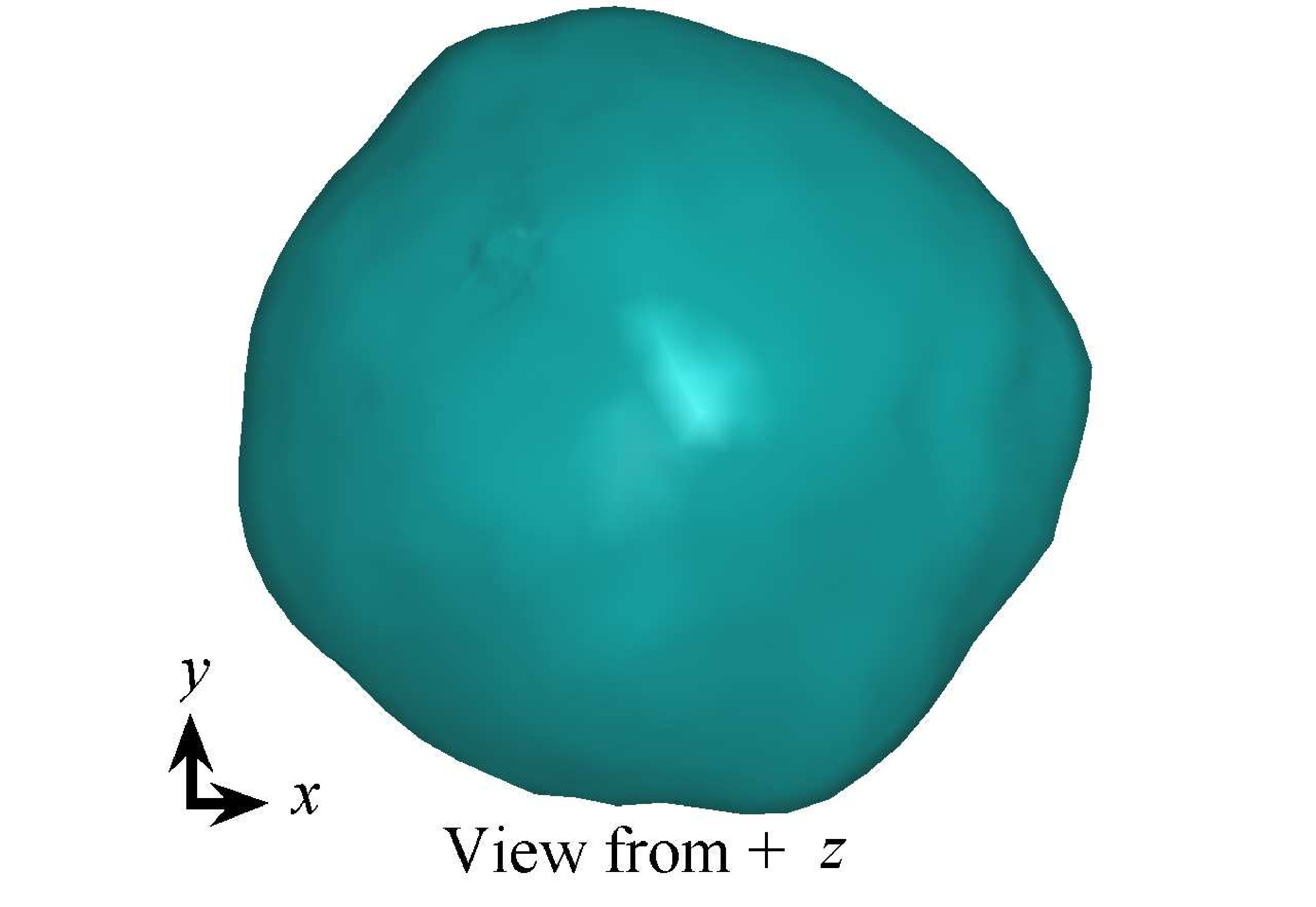}
 \includegraphics[width=41.5mm]{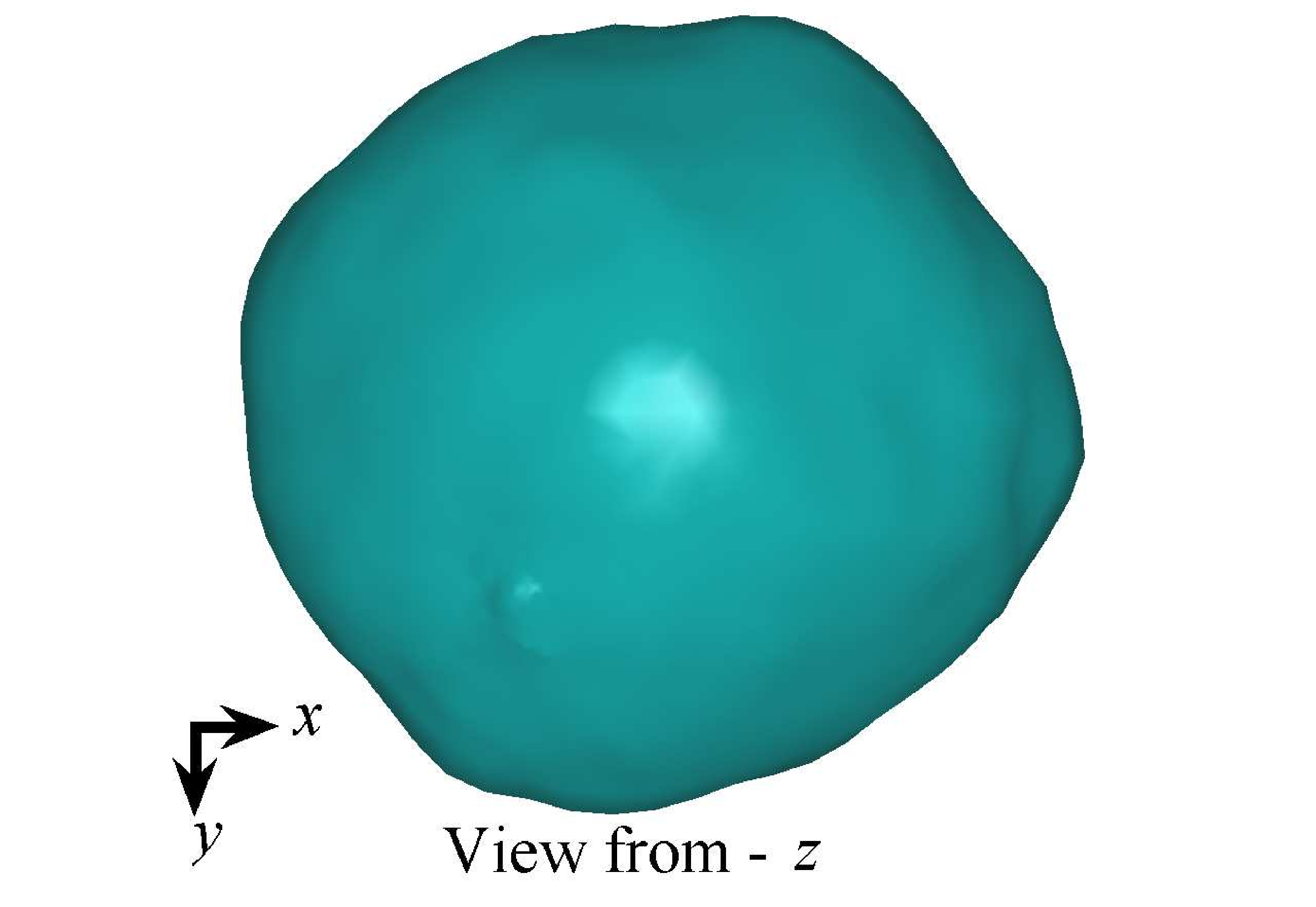}
 \caption{Polyhedral Model of Asteroid Bennu.}
\end{figure}

The contour plot of the radius of Bennu is shown in Fig. 2. The figure shows that there is a ridge near the equatorial region, which is quite common among asteroids whose size is small and rotation period is short in our solar system. Observations show that asteroids with sizes smaller than 125 km are likely to have a faster rotation velocity than expected, which may be caused by the YORP effect. Additionally, asteroids smaller than 50 km suffer from this effect much more. Therefore, the equatorial ridge is thought to be the result of the fast spin rate, which makes the surface material accumulate in the equatorial region. Most of the B-type asteroids are carbonaceous, with their surface covered in fine-grained regolith less than 1 cm in size. A thermal infrared study showed that there is not any extended emission of dust detectable, which indicates that the regolith only moves on the surface of Bennu \citep{b6}. Such a loose structure allows the shape of Bennu to be easily influenced by the dynamical environment in the proximity of the body.
\begin{figure*}
 \includegraphics[width=168mm]{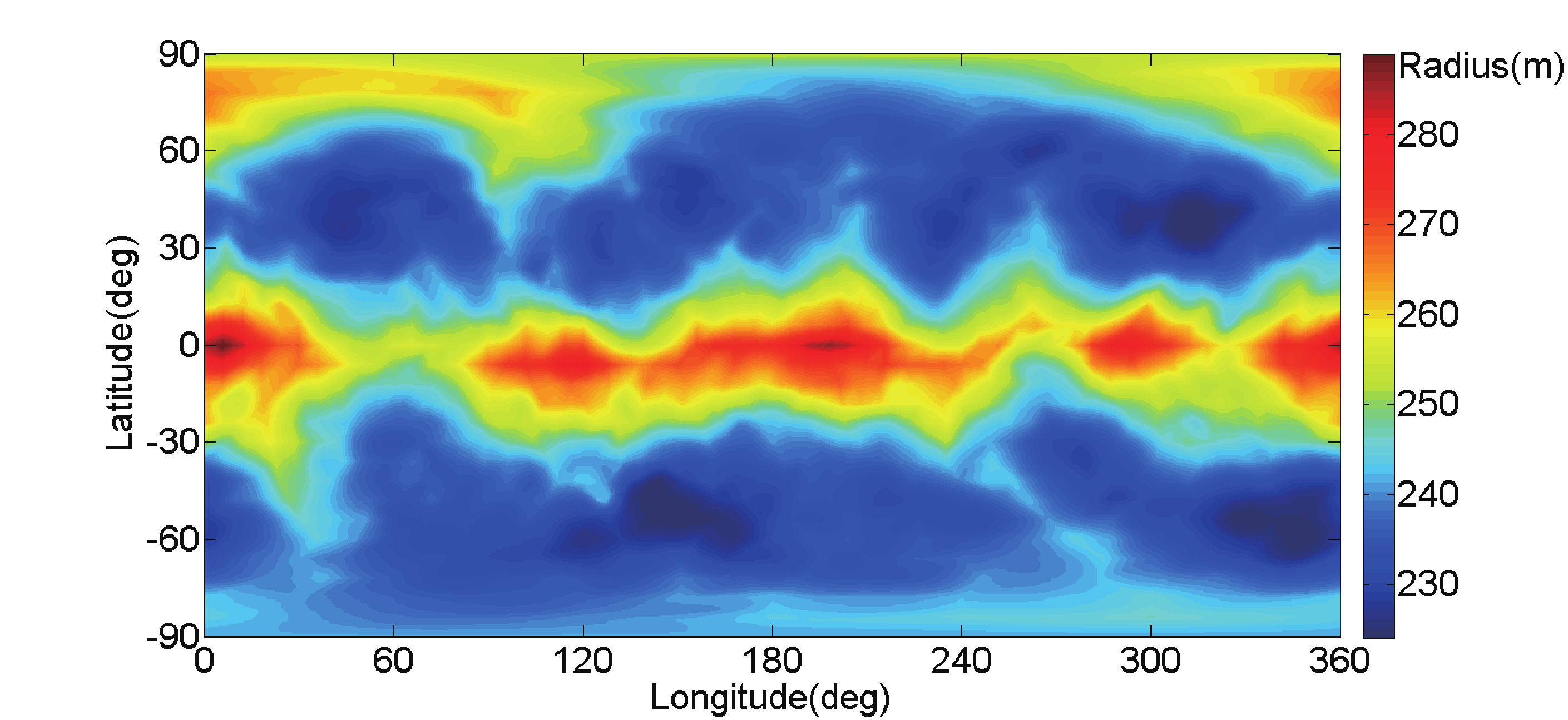}
 \caption{Radius contours of Asteroid Bennu.}
\end{figure*}

Generally, the asteroid is not homogeneous and consists of many different components. Thus, the density of the asteroid may vary in the body. However, in this case, we regard Bennu as a uniform body with the shape shown in Fig. 1 and a constant density. Considering the gravitational dynamics and YORP effect, the bulk density is estimated as $ 1260  \pm  70\ \rmn{kg m}^{-3}$, which is associated with a mass of $ (7.8 \pm 0.9) \times 10^{10}\ \rmn{kg} $. This result indicates a macroporosity in the range of $ 40 \pm  10 \%$, suggesting a rubble-pile internal structure for Bennu \citep{b3}. Because there are no other small bodies around Bennu to help predict the mass of the body, the estimation of the bulk density may be rough, and precious measurement is needed when the OSIRIS-REx spacecraft reaches this asteroid.

Using radar images and optical light curves, the rotation period of Bennu is derived as $4.29746 \pm  0.002\ \rmn{h}$, which is a rather fast spin speed that is supposed to still spin up. Principal-axis rotation is applied in this asteroid, with the pole orientation being $(-88^{\circ}, 45^{\circ})$ in the ecliptic coordinate system \citep{b12}. The body-fixed frame $O-xyz$ is defined as the origin located at Bennu¡¯s centre of mass, with the $x-$, $y-$, and $z-$axes corresponding to the principal axes of the smallest, intermediate, and largest moments of inertia, respectively. This definition indicates that the asteroid rotates along with the $z-$axis. The value of the principal moments of inertia are listed as follows:
\begin{equation}
   J_x=1.3749 \times 10^{15}\ \rmn{kg\,m}^2,
\end{equation}
\begin{equation}
   J_y=1.4285 \times 10^{15}\ \rmn{kg\,m}^2,
\end{equation}
\begin{equation}
   J_z=1.5421 \times 10^{15}\ \rmn{kg\,m}^2.
\end{equation}

There are little differences between the principal moments of inertia, indicating that Bennu has a mass distribution similar to a sphere with a little flattening. However, with the irregularity of the 3-D model, the orbital dynamics is much more complicated and quite different from that of the central gravitational field.

\section{EQUATIONS OF MOTION}

Studying the equation of motion is a fundamental way to understand the dynamical environment near an asteroid, which is quite different from the central gravitational field. In order to obtain a more general result, normalization is introduced. The shape model of Bennu is used, and numerical simulations are presented.

\subsection{Dynamical equations and nondimensionalization}

Considering the spacecraft as a massless particle orbiting the rotating Bennu, the equations of motion, which are time-varying in the inertial reference frame, can be easily written as autonomous differential equations in the body-fixed frame as follows:
\begin{equation}
   \ddot{\bmath{r}} + 2\bmath{\omega} \times \dot{\bmath{r}} + \bmath{\omega} \times (\bmath{\omega} \times \bmath{r})= \nabla{U(\bmath{r})},
\end{equation}
where $\bmath{r}$ is the body-fixed vector from the original point and $\bmath{\omega}$ is the angular velocity of Bennu. The body-fixed frame is described previously in Section 2. When applying the principal-axis rotation, $\bmath{\omega}$ is a constant with the magnitude $|\bmath{\omega}|=\omega$. $U(\bmath{r})$ is the gravitational potential, which is time-invariant and only related to the position of the field point. $\nabla$ is the gradient operator, and $\nabla{U(\bmath{r})}$ is the gravitational attraction. In this paper, $U(\bmath{r})$ and $\nabla{U(\bmath{r})}$ are calculated using a polyhedral method in the form of a summation \citep{b25}:
\begin{equation}
   U= \frac{1}{2} \rmn{G}\sigma \!\!\!\! \sum_{e \in edges} \!\!\!\! \bmath{r}_e \cdot \textbfss{E}_e \cdot \bmath{r}_e \cdot L_e - \frac{1}{2} \rmn{G}\sigma \!\!\!\! \sum_{f \in faces} \!\!\!\! \bmath{r}_f \cdot \textbfss{E}_f \cdot \bmath{r}_f \cdot \omega_f,
\end{equation}
\begin{equation}
   \nabla U=- \!\!\! \sum_{e \in edges} \!\!\!\!  \textbfss{E}_e \cdot \bmath{r}_e \cdot L_e + \rmn{G}\sigma \!\!\! \sum_{f \in faces} \!\!\!\!  \textbfss{E}_f \cdot \bmath{r}_f \cdot \omega_f,
\end{equation}
where $\rmn{G} = 6.67428 \times 10^{-11}\ \rmn{m}^{3}\rmn{kg}^{-1}\rmn{s}^{-2}$ represents the gravitational constant, $\sigma$ is the bulk density of the asteroid shape model, $\bmath{r}_a (a = e, f )$ is a body-fixed vector from the field point to any point on an edge (corresponding to subscript e) or a face (corresponding to subscript f), $L_e$ and $\omega_f$ are factors of integration that operate over the space between the field point and edges or faces, and $\textbfss{E}_e$ and $\textbfss{F}_f$ are dyads representing the geometric parameters of the edges and faces, which are defined in terms of face- and edge-normal vectors.

Considering the centrifugal force in the body-fixed frame, an efficient potential $V(\bmath{r})$ can be introduced as
\begin{equation}
   V(\bmath{r})=\frac{1}{2}(\bmath{\omega} \times \bmath{r})(\bmath{\omega} \times \bmath{r})+U(\bmath{r}).
\end{equation}
Combining Equations (4) and (7), a simpler form of the dynamical equations can be obtained as
\begin{equation}
   \ddot{\bmath{r}} + 2\bmath{\omega} \times \dot{\bmath{r}} = \nabla{V(\bmath{r})},
\end{equation}
which can also be written in scalar form as
\begin{equation}
   \ddot{x} - 2\omega \times \dot{y} = V_x=\omega^{2}x+U_x,
\end{equation}
\begin{equation}
   \ddot{y} + 2\omega \times \dot{x} = V_y=\omega^{2}y+U_y,
\end{equation}
\begin{equation}
   \ddot{z}  = V_z=U_z.
\end{equation}
One integral constant can be derived from the dynamical equations as
\begin{equation}
   J  = \frac{1}{2}\dot{\bmath{r}}\cdot\dot{\bmath{r}}-V(\bmath{r}),
\end{equation}
which is known as the Jacobi integral. Once given the position and velocity of the field point, the Jacobi integral will not change as the particle moves in the gravitational field. Moreover, the surface $J = -V(\bmath{r})$ is known as the zero velocity surface.

There are tens of thousands of asteroids in the solar system, and they range in size from metres to kilometres. The densities and rotation periods are also different among these small bodies. In order to obtain a more general result of the orbital dynamics near an asteroid, nondimensionalization was introduced. The equivalent radiu $r_0$ is the characteristic unit to scale the length, which is defined as follows:
\begin{equation}
   r_0=\sqrt[3]{3\rmn{Volume}/4\pi}.
\end{equation}
The physical meaning of the characteristic unit of length $r_0$ is the radius of a sphere that has the same volume as the asteroid. Additionally, the reciprocal of the angular velocity, $\omega^{-1}$, is the characteristic unit to scale the time.
For Bennu, the characteristic unit of length is
\begin{equation}
   [\rmn{L}]=r_0=\sqrt[3]{3\rmn{Volume}/4\pi}=245.88\rmn{m},
\end{equation}
and the characteristic unit of time is
\begin{equation}
   [\rmn{T}]=\omega^{-1}=2462.26\rmn{s}.
\end{equation}
Applying these nondimensionalizations to Equation (4), the nondimensionalized form of the dynamical equations is
\begin{equation}
   \ddot{\tilde{\bmath{r}}} + 2\tilde{\bmath{\omega}} \times \dot{\tilde{\bmath{r}}} + \tilde{\bmath{\omega}} \times (\tilde{\bmath{\omega}} \times \tilde{\bmath{r}})= \nabla{\tilde{U}(\tilde{\bmath{r}})},
\end{equation}
where quantities with the notation of $(\, \tilde{}\,)$ represent the scaled quantities, which are unitless. Thus, the scaled angular velocity $\tilde{\bmath{\omega}}$ has the very simple form of
\begin{equation}
   \tilde{\bmath{\omega}}=\left(\begin{array}{ccc} 0&0&1 \end{array}\right)^{\rmn{T}}.
\end{equation}
The gravitational attraction term $\nabla{\tilde{U}(\tilde{\bmath{r}})}$ also changes the form as
\begin{equation}
   \nabla{\tilde{U}(\tilde{\bmath{r}})}=\eta\left(- \!\!\! \sum_{e\in edges} \!\!\! \textbfss{E}_e\cdot\tilde{\bmath{r}}_e\cdot L_e + \!\!\! \sum_{f\in faces} \!\!\! \textbfss{F}_f\cdot\tilde{\bmath{r}}_f\cdot \omega_f\right),
\end{equation}
where $\eta$ is a dimensionless quantity that is defined as follows:
\begin{equation}
   \eta=\frac{\rmn{G}\sigma}{\omega^{2}},
\end{equation}
and the other quantities are as previously defined.
The nondimensionalized efficient potential $\tilde{V}(\tilde{\bmath{r}})$ and the nondimensionalized Jacobi constant $\tilde{J}$ can also be derived as
\begin{equation}
   \tilde{V}(\tilde{\bmath{r}})=\frac{1}{2}(\tilde{\bmath{\omega}} \times \tilde{\bmath{r}})(\tilde{\bmath{\omega}} \times \tilde{\bmath{r}})+\tilde{U}(\tilde{\bmath{r}}),
\end{equation}
\begin{equation}
   \tilde{J}  = \frac{1}{2}\dot{\tilde{\bmath{r}}}\cdot \dot{\tilde{\bmath{r}}}-\tilde{V}(\tilde{\bmath{r}}),
\end{equation}
Therefore, Equation (16) becomes
\begin{equation}
   \ddot{\tilde{\bmath{r}}} + 2\tilde{\bmath{\omega}} \times \dot{\tilde{\bmath{r}}} = \nabla{\tilde{V}(\tilde{\bmath{r}})}.
\end{equation}
In the nondimensionalized system, the volume of the asteroid is $ 4\pi/3 $ and the rotation period is $ 2\pi $ for all asteroids. The physical properties of an asteroid only affect the dimensionless quantity $\eta$ which appears in the gravitational potential and attraction terms. Equation (22) shows that only the right-hand side is different among the different asteroids. There are two main factors that lead to this difference. One is the shape of the asteroid, and the other is the dimensionless quantity $ \eta $. This result can be easily seen from Equation (18). Using these nondimensionalized equations of motion, the results can be translated to a small asteroid to a big one with a similar shape.

The bulk density $\sigma$ and the angular velocity $\omega$ of Bennu are $1260 \pm  70\ \rmn{kg m} ^{-3}$ and $4.0613 \times 10 ^{-4}\ \rmn{rad s} ^{-1}$ respectively, so the dimensionless quantity $ \eta$ can be obtained as
\begin{equation}
   \eta=\frac{\rmn{G}\sigma}{\omega^{2}}=0.5095.
\end{equation}
When applying the characteristic units of length and time to the nondimensionalized system, the physical quantities in SI units can be obtained.

\subsection{Equilibrium points and their stability}

The equilibrium points are special orbits that remain relatively static in the body-fixed frame and are characterized by zero acceleration and zero velocity. According to Equation (22), the equilibrium points must satisfy the following equations:
\begin{equation}
   \nabla\tilde{V}(\tilde{\bmath{r}})=0.
\end{equation}
By solving Equation (24), the equilibrium points can be obtained numerically. The results are shown in Table 1. There are nine equilibrium points in the gravitational field of Bennu. Eight of them are outside the small body, and one of them is inside. The equilibrium points are not in the $x-y$ plane because of the asymmetry of Bennu's shape. However, they are very close to the $x-y$ plane with a small $z-$axis component. The zero velocity curves and equilibrium points are plotted in the $x-y$ plane in Fig. 3.
\begin{figure}
 \includegraphics[width=84mm]{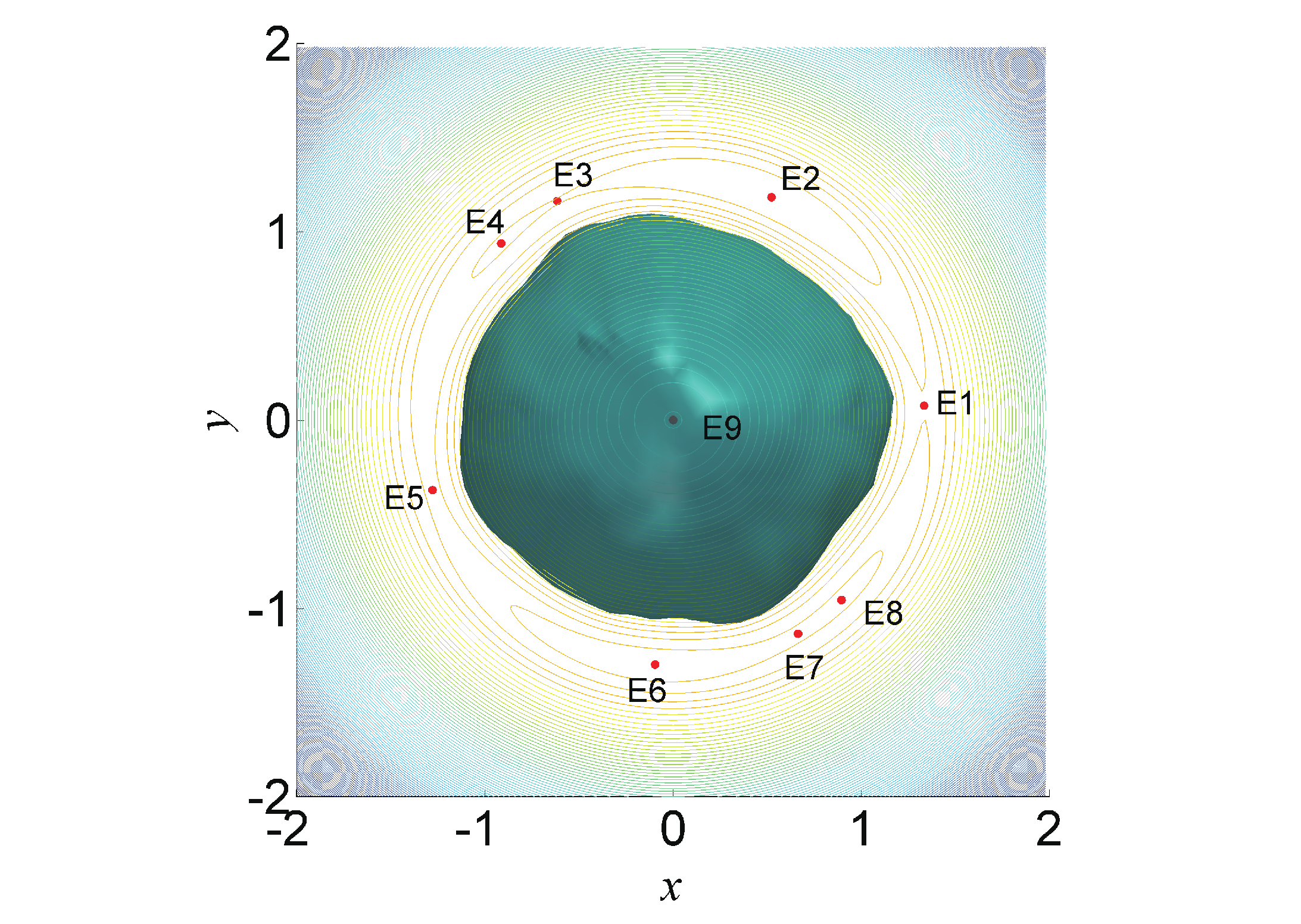}
 \caption{The Equilibrium Points and Zero Velocity Curves of Bennu.}
\end{figure}

The Jacobi integral can be calculated using Equation (21) by applying zero velocity, and the values of each equilibrium point are also shown in Table 1.

\begin{table}
 \caption{Locations of Equilibrium Points of Asteroid Bennu.}
 \label{symbols}
 \begin{tabular}{@{}lrrrc }
  \hline
   \makecell{Equilibrium\\ Points}  & $x$ & $y$
        & $z$
        & \makecell{Jacobi\\ Constant} \\
  \hline
  E1    &$1.333$ & $0.076$ & $-0.013$ & $-2.53545$\\
  E2    &$0.522$ & $1.184$ & $-0.010$ & $-2.49552$\\
  E3    &$-0.616$ & $1.164$ & $-0.032$ & $-2.51460$\\
  E4    &$-0.914$ & $0.938$ & $-0.030$ & $-2.51299$\\
  E5    &$-1.278$ & $-0.372$ & $-0.009$ & $-2.53181$\\
  E6    &$-0.096$ & $-1.298$ & $0.001$ & $-2.50264$\\
  E7    &$0.664$ & $-1.135$ & $-0.008$ & $-2.51136$\\
  E8    &$0.895$ & $-0.955$ & $-0.012$ & $-2.51075$\\
  E9    &$0.000$ & $0.001$ & $0.000$ & $-3.19577$\\
  \hline
 \end{tabular}
\end{table}

By setting $\tilde{\bmath{r}}=0$ in Equation (21), we can get the 3-D zero velocity surface in the potential field of Bennu. For each certain value of Jacobi constant $C$, there are an allowable region, which has $¨CV(\tilde{\bmath{r}}) \leq C$, and a forbidden region, which has $¨CV(\tilde{\bmath{r}})>C$, separating by zero velocity surface. The zero velocity surfaces with different values of Jacobi constant $C$ are shown in Fig. 4. The inner and outer allowable areas connect to each other if the Jacobi constant $C$ increases. The allowable area of motion enlarges and the forbidden area shrinks. If the Jacobi constant $C$ continue to increase, the whole space will be allowable area of motion.
\begin{figure*}
 \includegraphics[width=57mm]{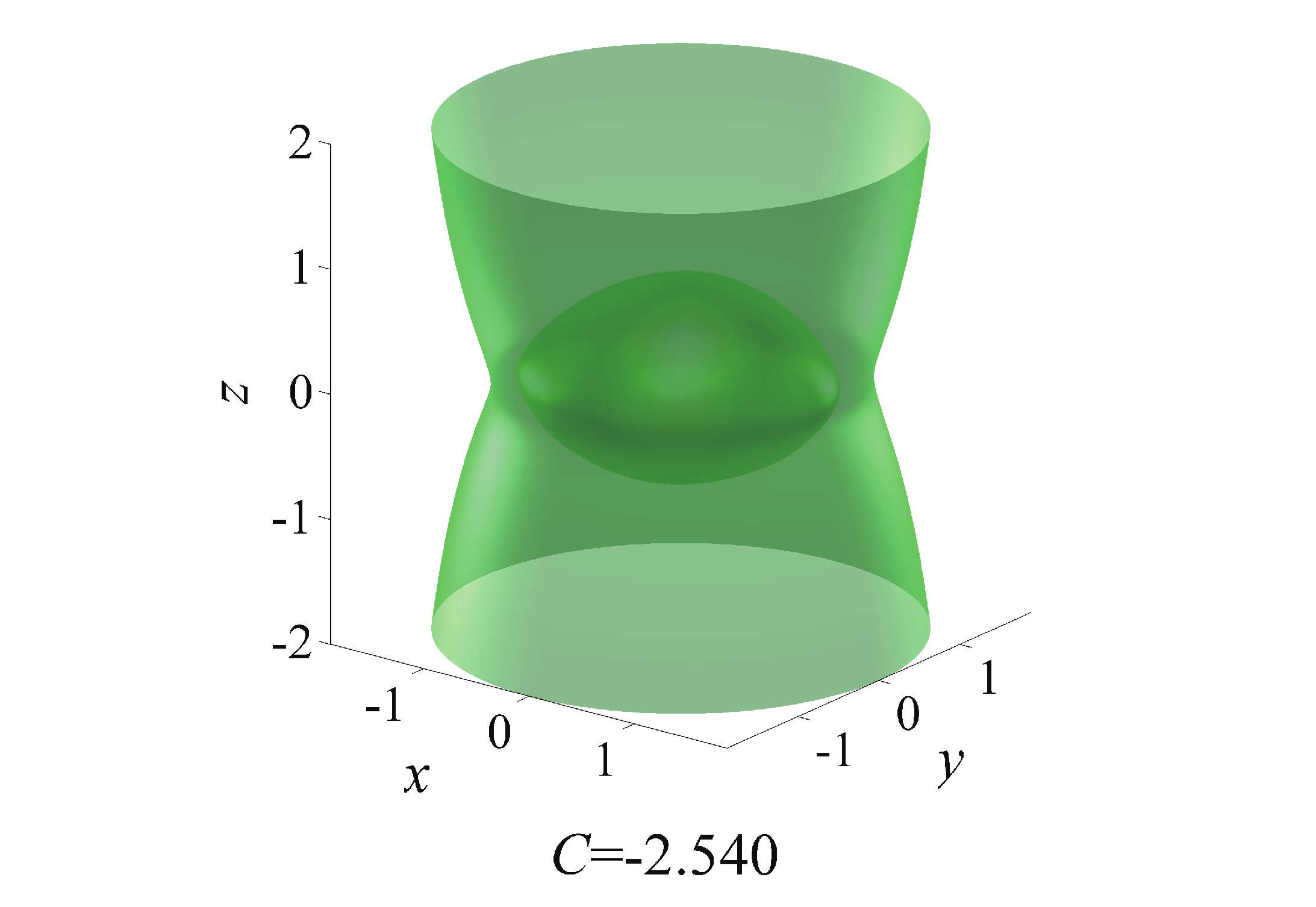}
 \vspace{10pt}
 \includegraphics[width=57mm]{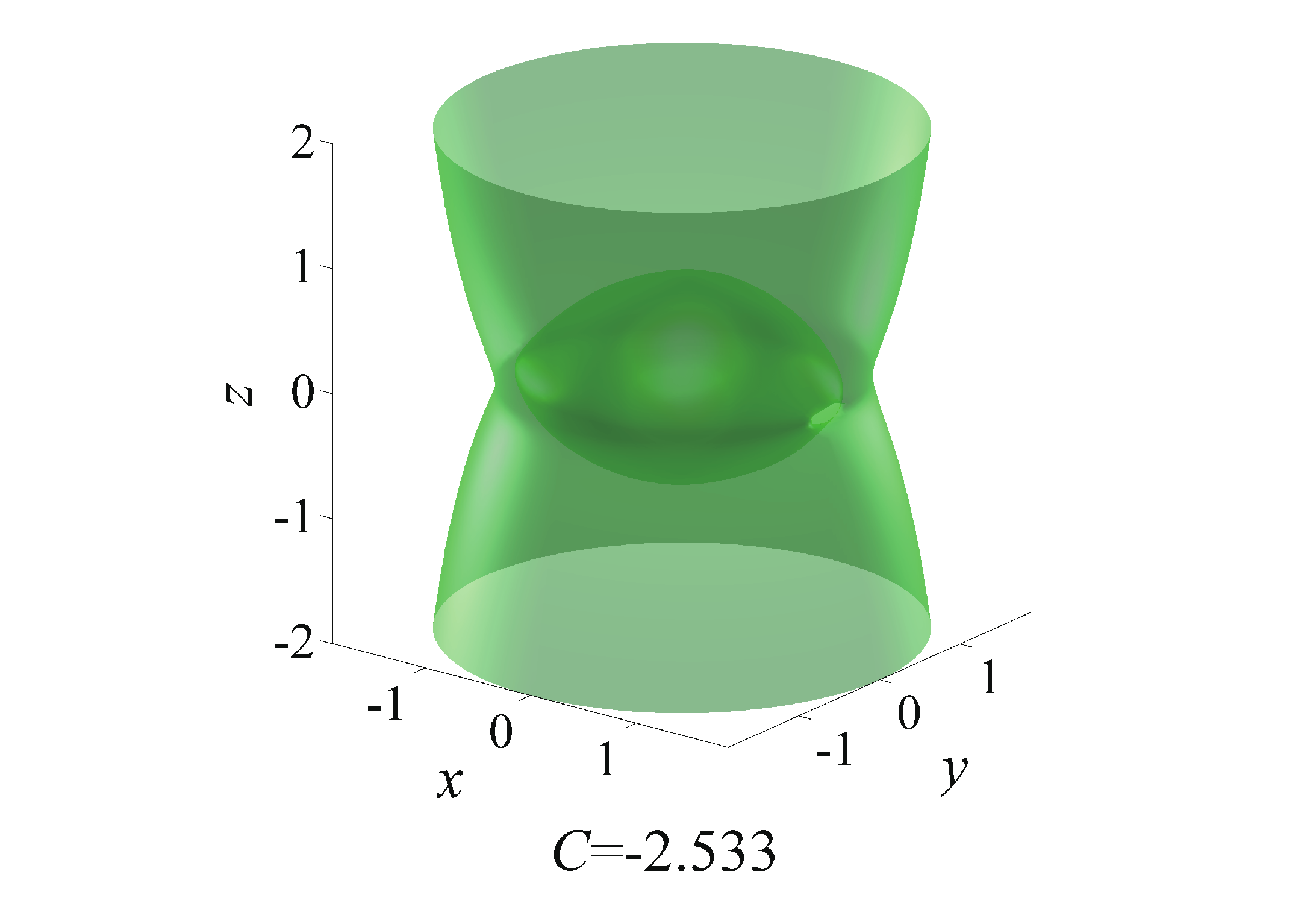}
 \includegraphics[width=57mm]{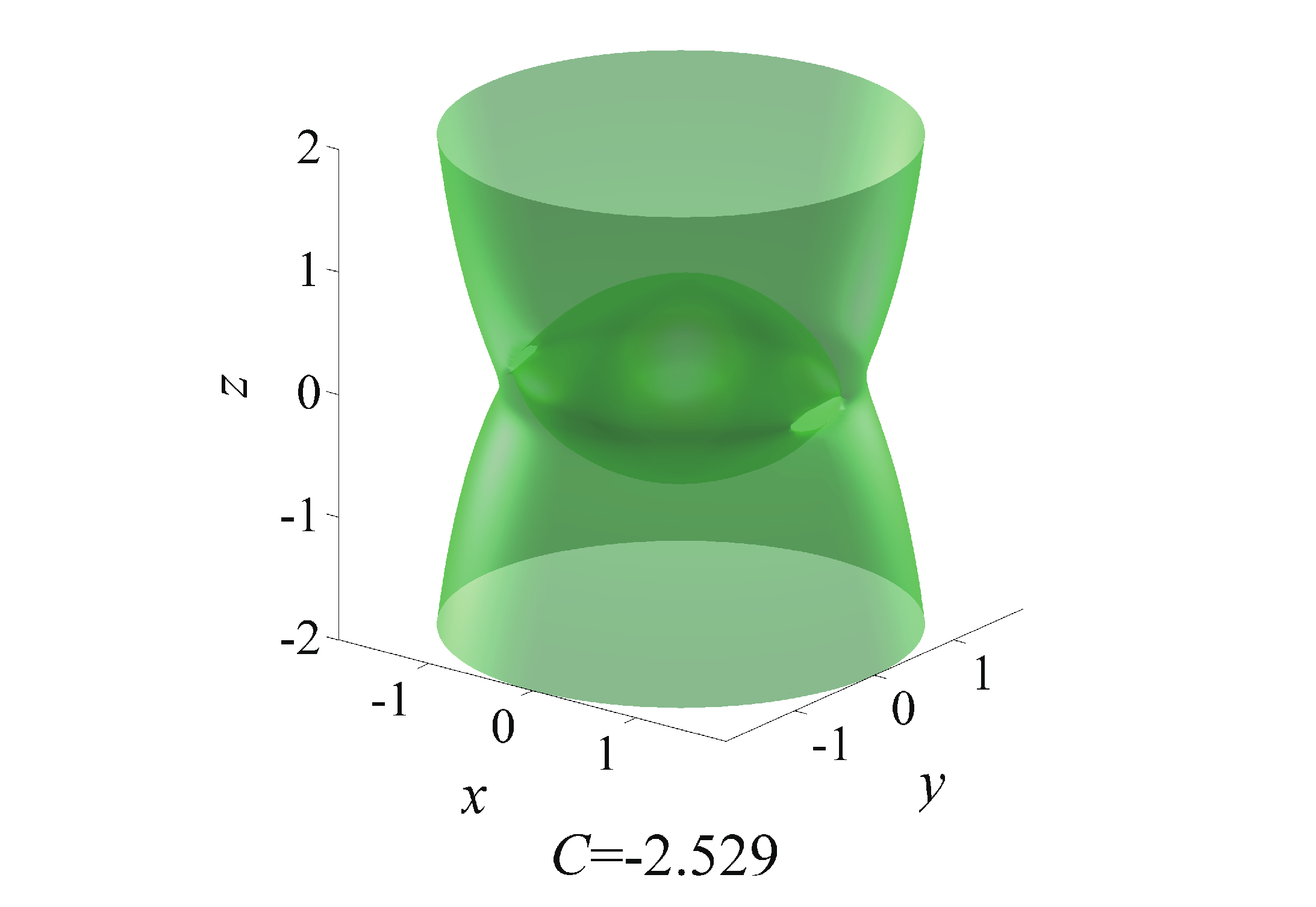}
 \includegraphics[width=57mm]{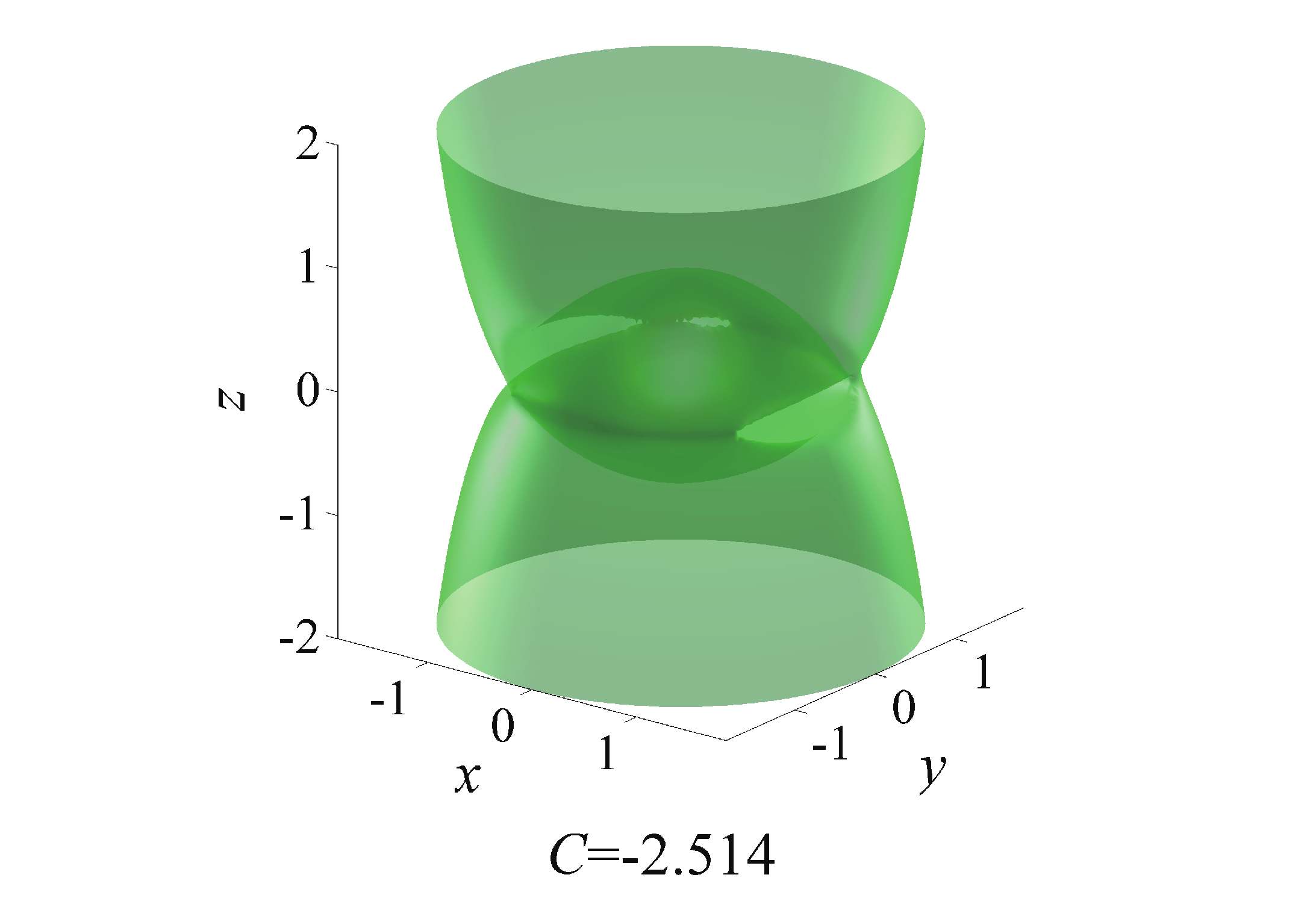}
 \vspace{10pt}
 \includegraphics[width=57mm]{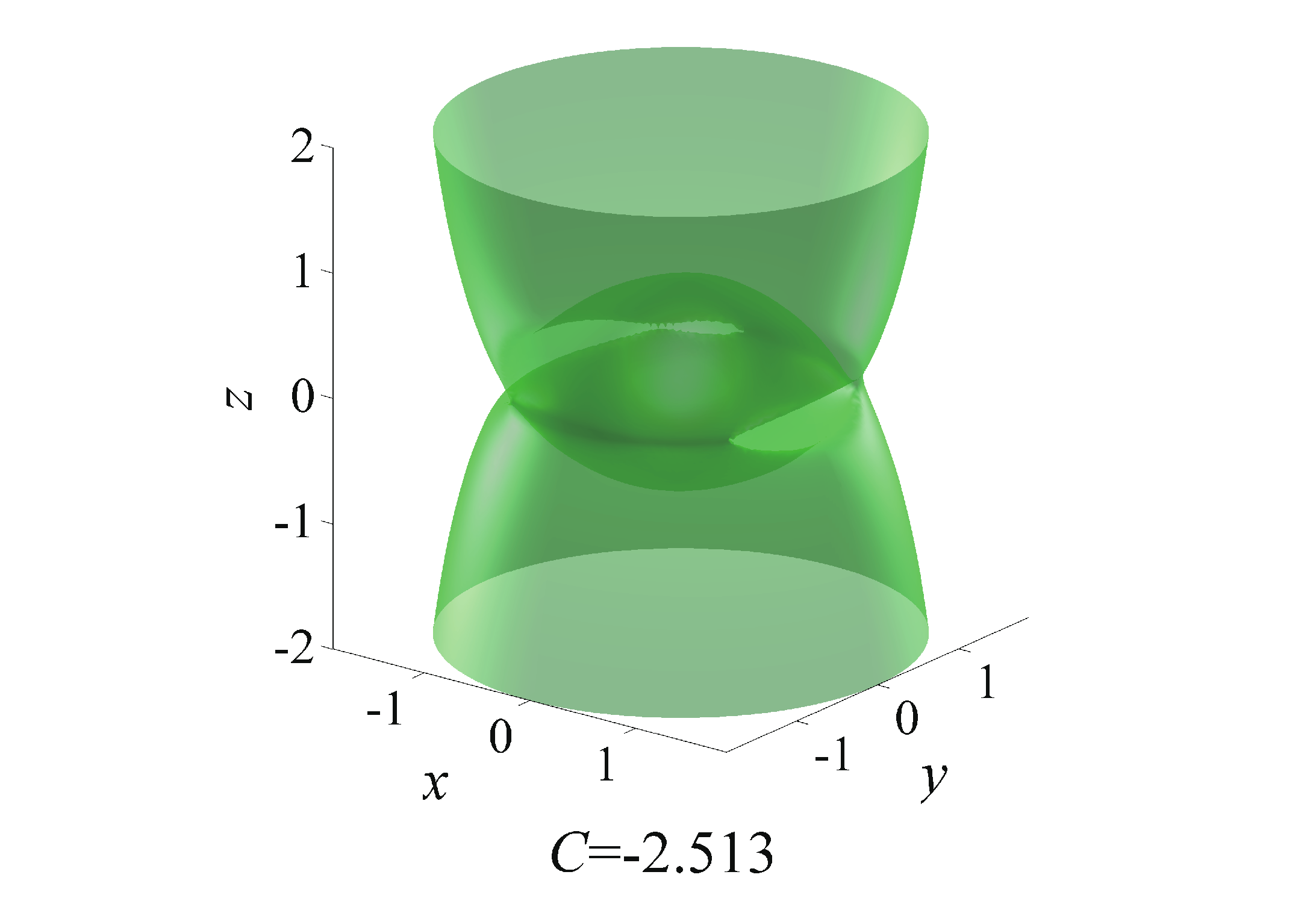}
 \includegraphics[width=57mm]{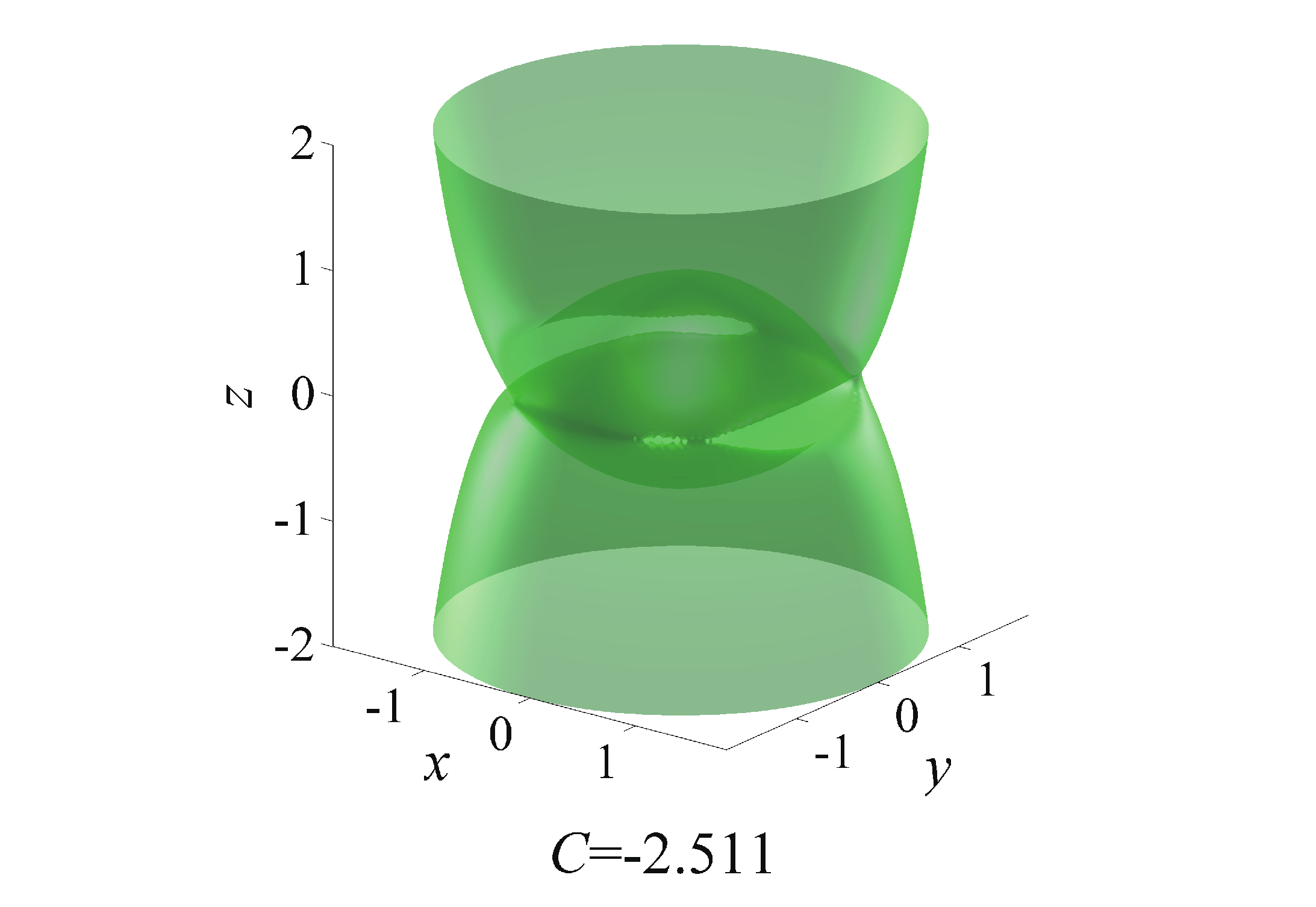}
 \includegraphics[width=57mm]{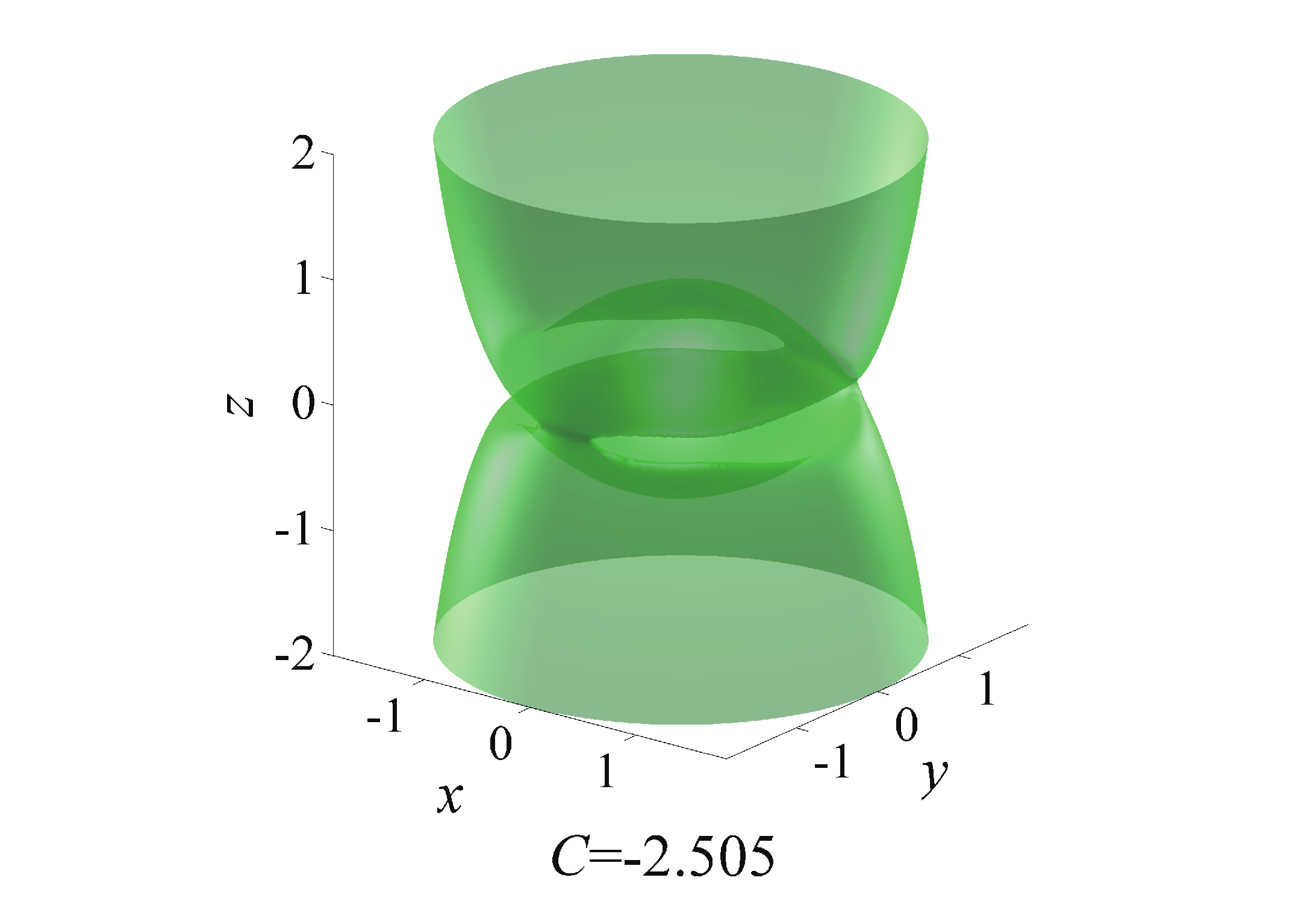}
 \includegraphics[width=57mm]{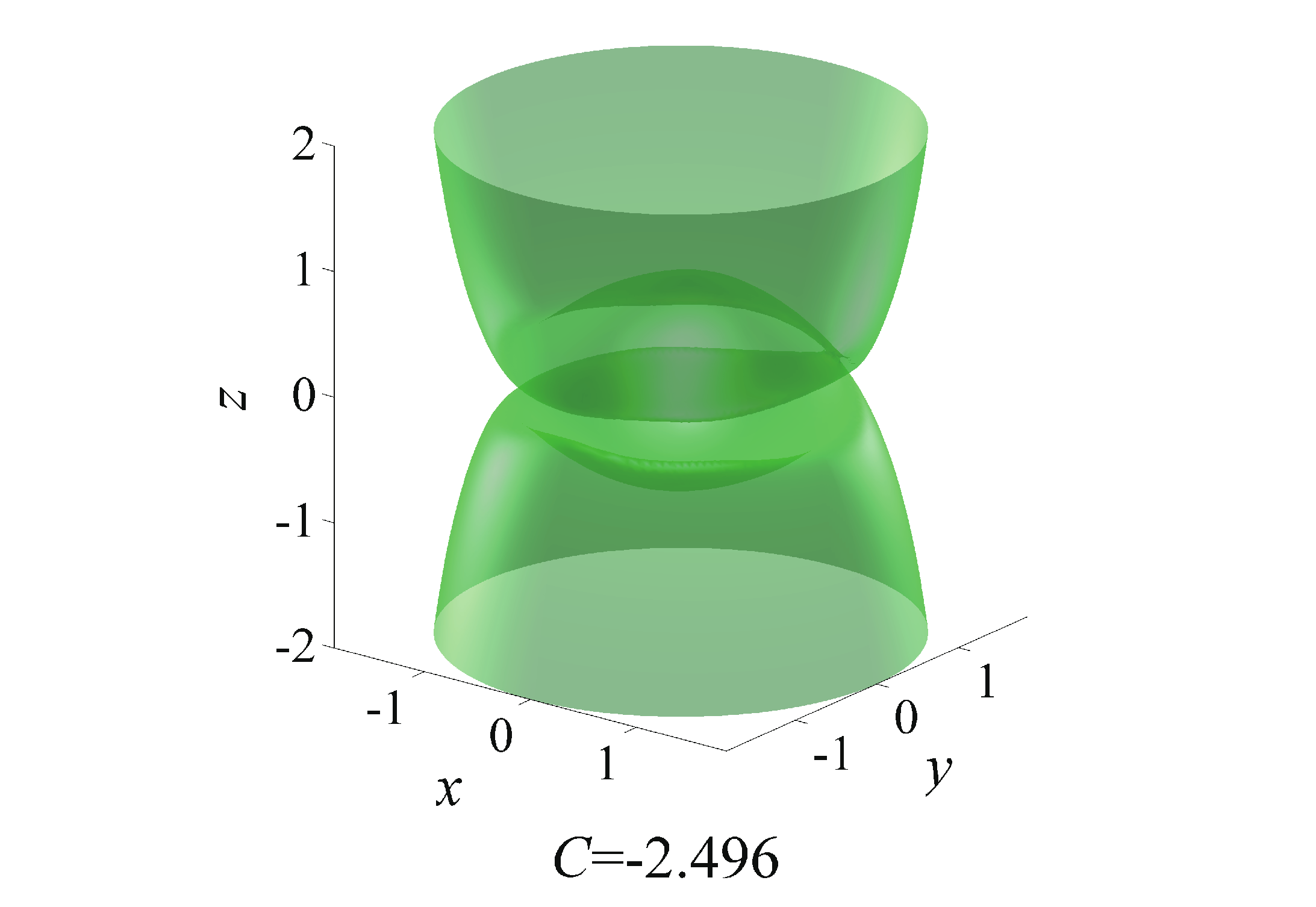}
 \includegraphics[width=57mm]{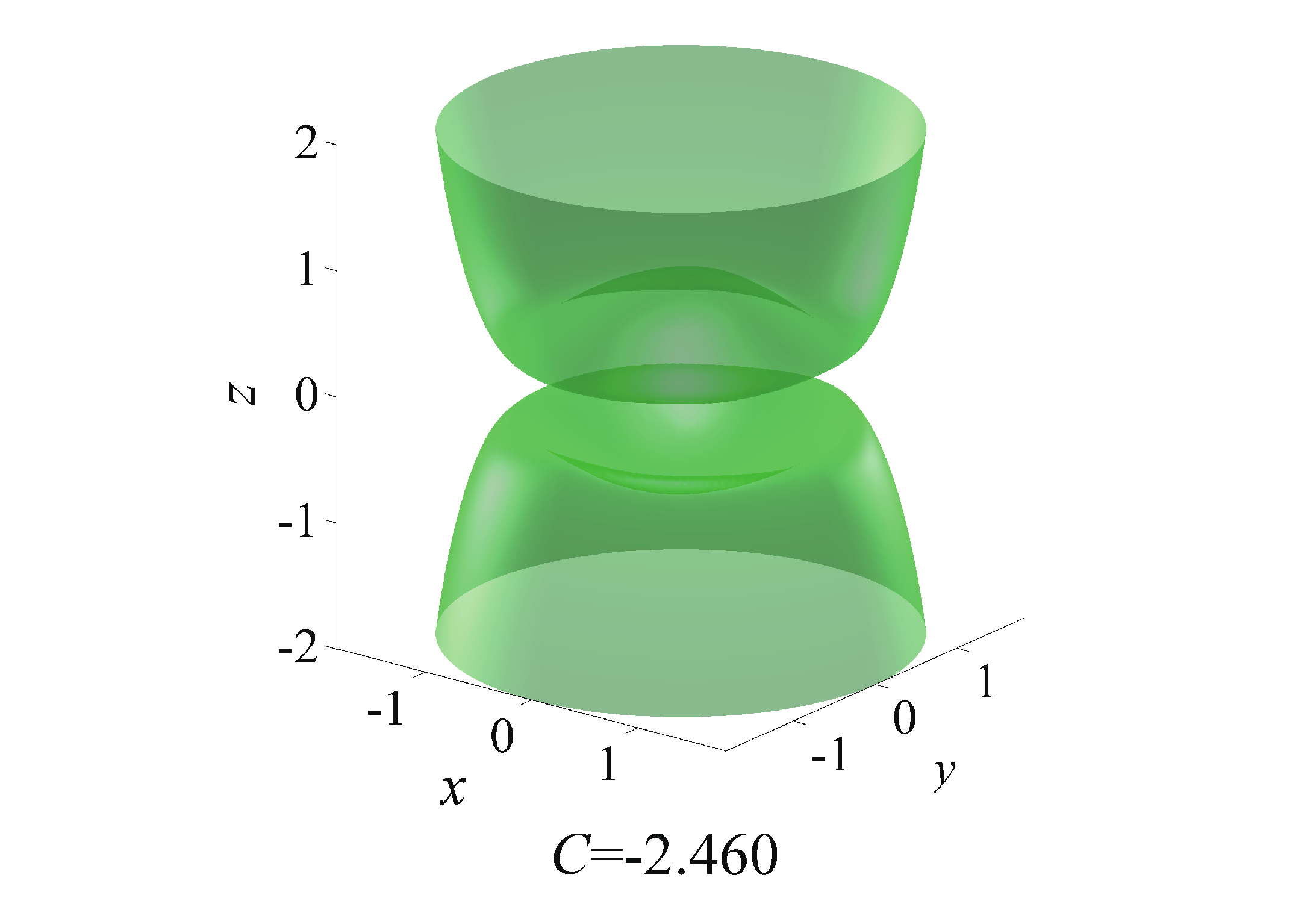}
 \caption{Evolution of the Zero Velocity Surfaces with Different Values of Jacobi Constant.}
\end{figure*}

In the inertial frame, these equilibrium points represent period orbits that have the same period as Bennu's rotation and are very helpful in an asteroid exploration mission. Before using the equilibrium points, their stability must be studied first. The linearization method is used the most and is applied in this research.

Let us denote the coordinates of the equilibrium point as $ (\begin{array}{ccc} \tilde{x}_e ,&\!\!\!\tilde{y}_e ,&\!\!\!\tilde{z}_e \end{array})^{\rmn{T}}$, and the perturbation variables are $ (\begin{array}{ccc} \xi ,&\!\!\!\zeta ,&\!\!\!\varsigma \end{array})^{\rmn{T}}$, where
\begin{equation}
   \xi=\tilde{x}-\tilde{x}_e,
\end{equation}
\begin{equation}
   \zeta=\tilde{y}-\tilde{y}_e,
\end{equation}
\begin{equation}
   \varsigma=\tilde{z}-\tilde{z}_e.
\end{equation}
Using the linearization method, the perturbation equations of motion with respect to an equilibrium point can be written as
\begin{equation}
   \dot{\bmath{\varepsilon}}=\textbfss{A}\bmath{\varepsilon},
\end{equation}
where $ \bmath{\varepsilon}=(\begin{array}{cccccc} \xi & \zeta & \varsigma & \dot{\xi} & \dot{\zeta} & \dot{\varsigma}\end{array})^{\rmn{T}}$ are the state variables of the linearized system, and $ \textbfss{A}$ is the Jacobi matrix of the linearized system which is
\begin{equation}
   \textbfss{A}=\left[ \begin{array}{cc} \textbfss{0} & \textbfss{I} \\ \nabla^{2}\tilde{V}& -2\hat{\bmath{\omega}} \end{array} \right],
\end{equation}
where $ \textbfss{I}$ is a unit matrix and
\begin{equation}
   \nabla^{2}\tilde{V}=\left[ \begin{array}{ccc} \tilde{V}_{xx} &\tilde{V}_{xy} &\tilde{V}_{xz} \\ \tilde{V}_{yx} & \tilde{V}_{yy} & \tilde{V}_{yz} \\ \tilde{V}_{zx} & \tilde{V}_{zy} & \tilde{V}_{zz} \end{array} \right],
\end{equation}
\begin{equation}
   \hat{\bmath{\omega}}=\left[ \begin{array}{ccc} 0&-1&0\\1&0&0\\0&0&0 \end{array} \right].
\end{equation}
Equation (28) can also be written in scalar form as
\begin{equation}
   \ddot{\xi}-2\dot{\zeta}-\tilde{V}_{xx}\xi-\tilde{V}_{xy}\zeta-\tilde{V}_{xz}\varsigma=0,
\end{equation}
\begin{equation}
   \ddot{\zeta}+2\dot{\xi}-\tilde{V}_{yx}\xi-\tilde{V}_{yy}\zeta-\tilde{V}_{yz}\varsigma=0,
\end{equation}
\begin{equation}
   \ddot{\varsigma}-\tilde{V}_{zx}\xi-\tilde{V}_{zy}\zeta-\tilde{V}_{zz}\varsigma=0.
\end{equation}
The characteristic equations of the linearized system relating to an equilibrium point can be obtained using Equations (32)-(34):
\begin{equation}
\begin{aligned}
   &\lambda^{6}-(\tilde{V}_{xx}+\tilde{V}_{yy}+\tilde{V}_{zz}+4)\lambda^{4}- (\tilde{V}_{xx}\tilde{V}_{yy}+\tilde{V}_{yy}\tilde{V}_{zz}\\
  & + \tilde{V}_{zz}\tilde{V}_{xx}  -\tilde{V}^{2}_{xy} -\tilde{V}^{2}_{yz}- \tilde{V}^{2}_{zx} + 4\tilde{V}_{zz})\lambda^{2}-( \tilde{V}_{xx}\tilde{V}_{yy}\tilde{V}_{zz} \\
  &+ 2\tilde{V}_{xy}\tilde{V}_{yz}\tilde{V}_{zx} - \tilde{V}_{xx}\tilde{V}^{2}_{yz}- \tilde{V}_{yy}\tilde{V}^{2}_{xz}- \tilde{V}_{zz}\tilde{V}^{2}_{xy})=0.
\end{aligned}
\end{equation}
By solving Equation (35), the eigenvalues of the Jacobi matrix can be obtained and are shown in Table 2.

\begin{table*}
 \centering
 \begin{minipage}{140mm}
 \caption{Eigenvalues of Jacobi Matrix of Equilibrium Points.}
 \begin{tabular}{@{}lrrrrrrc}
  \hline
  {Eigenvalues} & $\lambda_1$ & $\lambda_2$ &$\lambda_3$ & $\lambda_4$ & $\lambda_5$ & $\lambda_6$ & Topological structure \\
  \hline
  E1    & 1.14i & -1.14i & 1.05i & -1.05i & 0.63 & -0.63 & Case2 \\
  E2    & 1.06i & -1.06i & 0.10+0.67i & 0.10-0.67i & -0.10+0.67i & -0.10-0.67i & Case5 \\
  E3    & 1.11i & -1.11i & 0.98i & -0.98i & 0.46 & -0.46 & Case2 \\
  E4    & 1.09i & -1.09i & 0.11+0.64i & 0.11-0.64i & -0.11+0.64i & -0.11-0.64i & Case5 \\
  E5    & 1.13i & -1.13i & 1.00i & -1.00i & 0.54 & -0.54 & Case2 \\
  E6    & 1.07i & -1.07i & 0.09+0.66i & 0.09-0.66i & -0.09+0.66i & -0.09-0.66i & Case5 \\
  E7    & 1.11i & -1.11i & 0.95i & -0.95i & 0.39 & -0.39 & Case2 \\
  E8    & 1.09i & -1.09i & 0.76i & -0.76i & 0.48i & -0.48i & Case1 \\
  E9    & 2.43i & -2.43i & 1.51i & -1.51i & 0.43i & -0.43i & Case1 \\
  \hline
 \end{tabular}
 \end{minipage}
\end{table*}

The stability of these equilibrium points can be judged by the corresponding eigenvalues. If the eigenvalue has a positive real part, then the equilibrium point is unstable. Otherwise, the equilibrium point is linear stable. The eigenvalues of the Jacobi matrix can also indicate the topological structure of the equilibrium points. According to \citet{b9}, the topological structure of a non-degenerate and non-resonance equilibrium point can be classified into five cases, which are shown in Table 3. Generally, there are six eigenvalues in the linear system. Equation (35) shows that only the even powers of $\lambda$ exist. Thus, if $\lambda$  is one of the eigenvalues, then $-\lambda$  and $\bar{\lambda}$  are both eigenvalues of the linear system. The six eigenvalues must be a combination of the following situations: 1) a pair of opposite real roots, i.e., $\pm \alpha$; 2) a pair of conjugate imaginary roots, i.e., $ \pm \rmn{i}\beta$; 3) two pairs of opposite and conjugate complex roots, i.e., $\pm \sigma \pm \rmn{i}\tau$, where $\alpha$, $\beta$, $\sigma$, and $\tau$ are any positive real quantities. All possible combinations are shown in Table 3. In Case 1, the dimension of the central manifold is six, indicating there are three families of the period orbit near the equilibrium point. In Case 2, the dimension of both the stable and unstable manifolds is one, and the dimension of the central manifold is four, indicating there are two families of the period orbit near the equilibrium point. In Case 3, the dimension of the stable manifold, unstable manifold, and central manifold is two, indicating there is one family of the period orbit near the equilibrium point. In Case 4, the dimension of both the stable and unstable manifolds is three, and the dimension of the central manifold is zero, indicating there is no period orbit near the equilibrium point. In Case 5, the dimension of both the stable and unstable manifolds is two, and the dimension of the central manifold is also two, indicating there is one family of the period orbit near the equilibrium point. The stability of the equilibrium point is also shown in Table 3. The topological structure of the equilibrium points of Bennu can be distinguished based on the eigenvalues. The results are also shown in Table 2.

\begin{table*}
 \centering
 \begin{minipage}{140mm}
 \caption{Topological Classification of Non-degenerate and Non-resonance Equilibria \citep{b9}.}
 \begin{tabular}{@{}lccc}
  \hline
  {Case} & Eigenvalues & Number of period families
        & Stability \\
  \hline
  Case1    &$ \pm \rmn{i}\beta ( \beta_j \in \bld{R}^+ ; j= 1, 2, 3) $ & $3$ & LS\\
  Case2    &$ \pm \alpha_j (\alpha_j \in \bld{R}^{+} ; j=1), \pm \rmn{i}\beta ( \beta_j \in \bld{R}^{+} ; j= 1, 2) $ & $2$ & U\\
  Case3    &$ \pm \alpha_j (\alpha_j \in \bld{R}^{+} ; j=1,2), \pm \rmn{i}\beta ( \beta_j \in \bld{R}^{+} ; j= 1) $ & $2$ & U\\
  Case4a     &$ \pm \alpha_j (\alpha_j \in \bld{R}^{+} ; j=1, \pm \sigma \pm \rmn{i}\tau ( \sigma, \tau \in \bld{R}^{+} ) $ & $1$ & U\\
  Case4b    &$ \pm \alpha_j ( \alpha_j \in \bld{R}^+ ; j= 1, 2, 3 $ & $0$ & U\\
  Case5    &$ \pm \sigma \pm \rmn{i}\tau ( \sigma, \tau \in \bld{R}^{+} ) , \pm \rmn{i}\beta ( \beta_j \in \bld{R}^+ ; j= 1) $ & $1$ & U\\
  \hline
 \end{tabular}
 \end{minipage}
\end{table*}

The topological structure near the equilibrium point can be presented by the eigenvalues of the Jacobi matrix. According to Table 2, E1, E3, E5, and E7 have one pair of one-dimensional unstable/stable manifolds and two two-dimensional centre manifolds. Fig. 5 shows the results of E1. The topological structures of E3, E5, and E7 are similar to that of E1 so are not presented here. E2, E4, and E6 have one pair of two-dimensional unstable/stable manifolds and one two-dimensional central manifold. Fig. 6 shows the results of E2. The topological structures of E4 and E6 are similar to that of E2 so are not shown here. E8 and E9 have three two-dimensional central manifolds. The results of E8 are shown in Fig. 7, and E9 has similar results.

\begin{figure}
 \includegraphics[width=84mm]{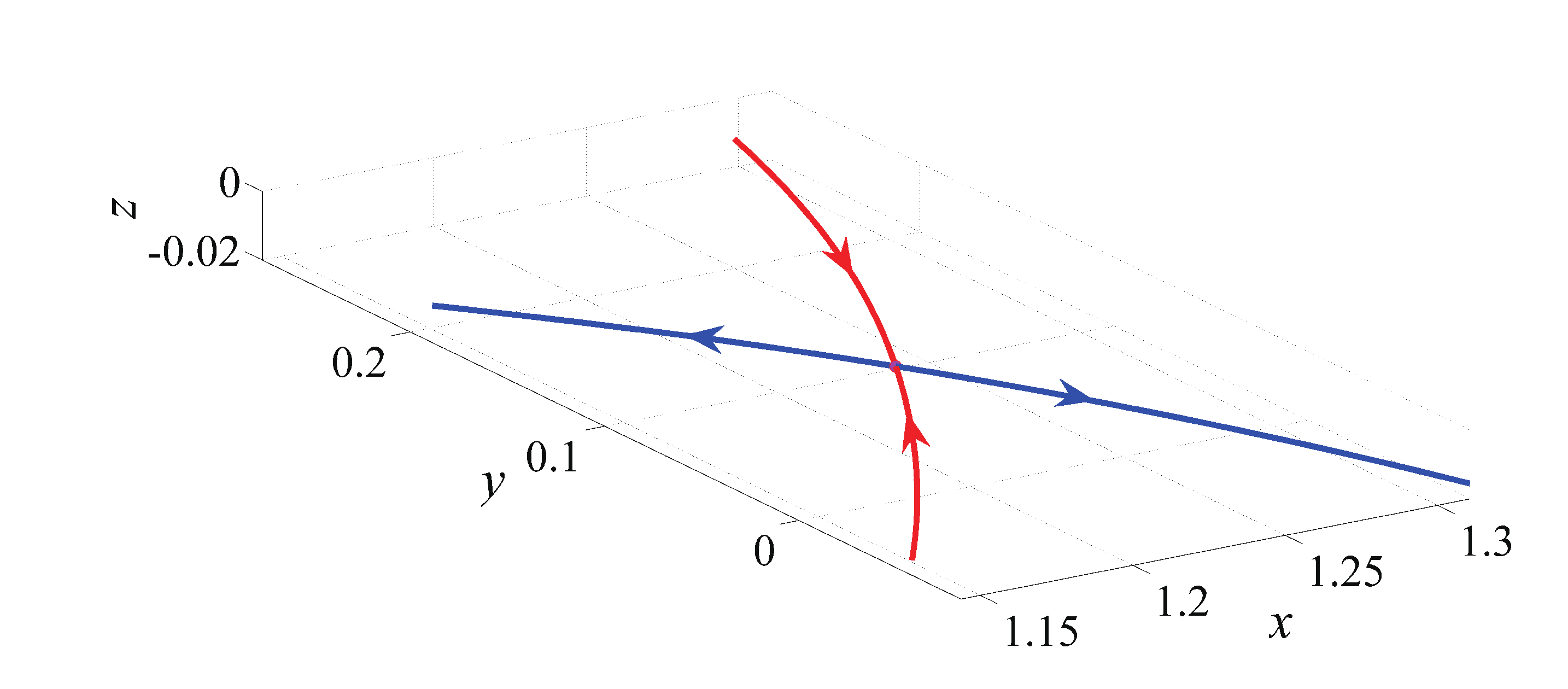}
 a. A pair of unstable/stable manifolds of E1. The red line represents the stable manifolds and the blue line represents the unstable manifolds.\\
 \includegraphics[width=84mm]{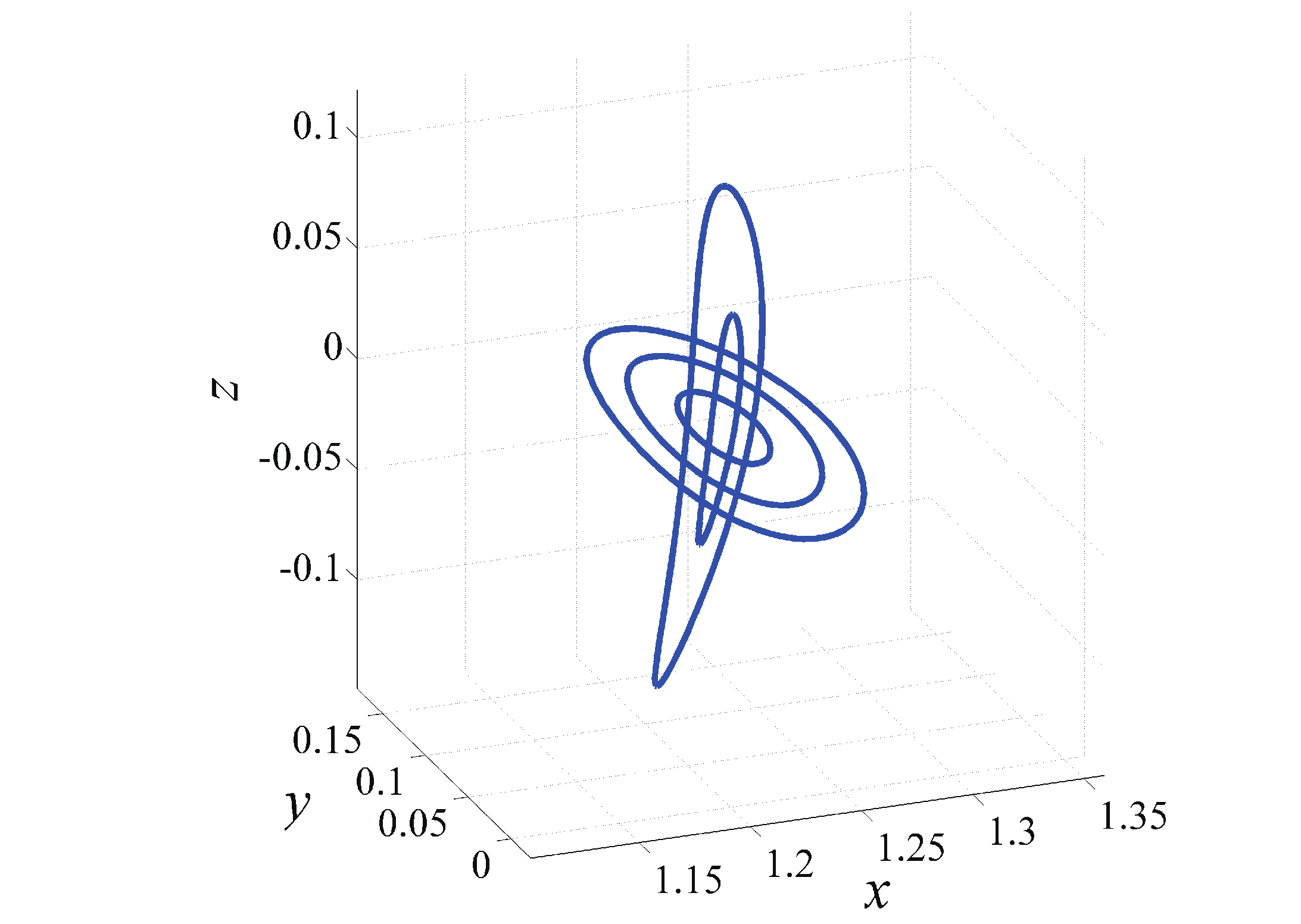}
 b. Two 2-dimensional center manifolds of E1.\\
 \caption{Topological structure of manifolds of E1.}
\end{figure}
\begin{figure}
 \includegraphics[width=84mm]{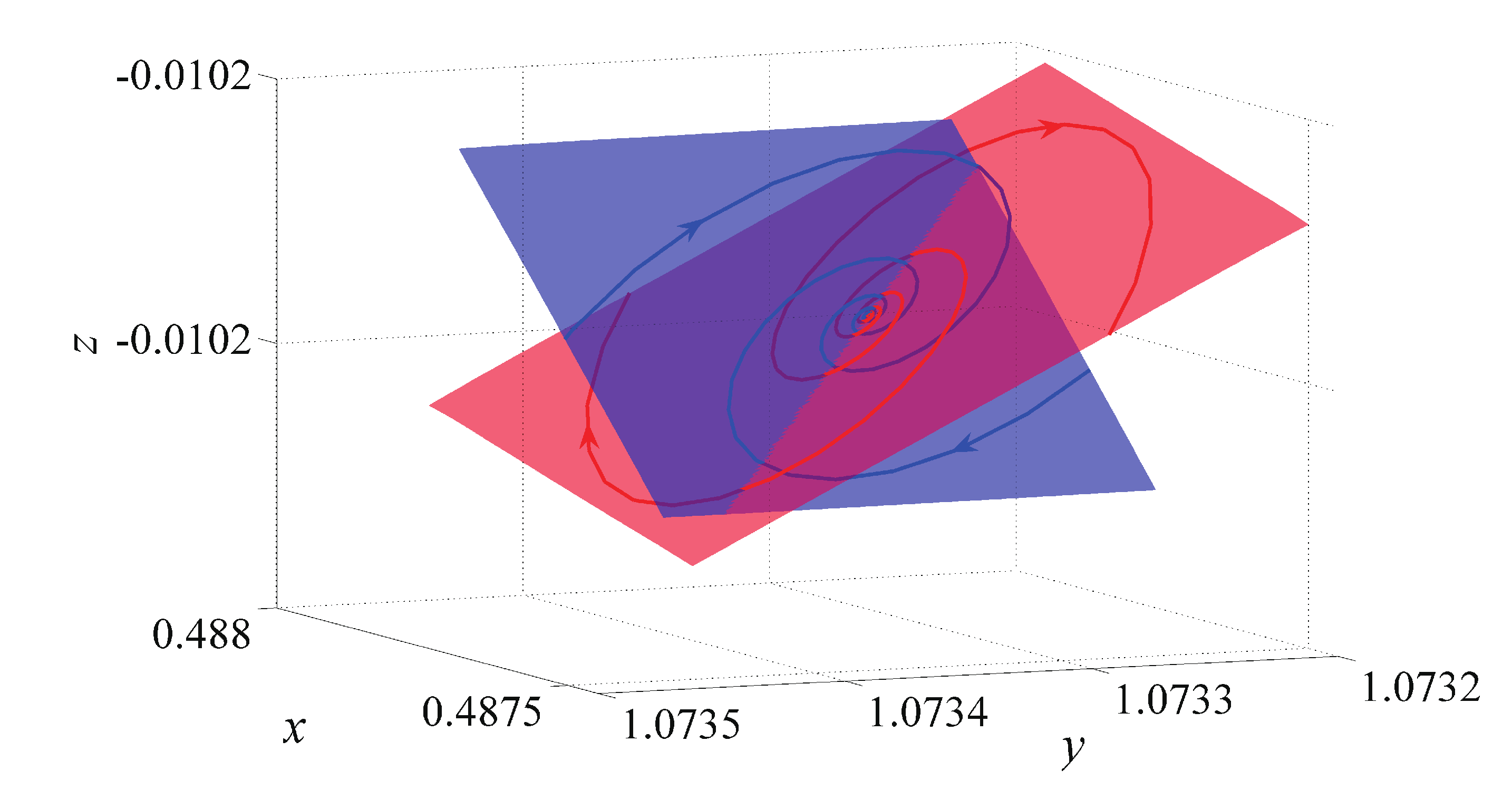}
 a. Two pairs of unstable/stable manifolds of E2. The red face represents the stable manifolds and the blue face represents the unstable manifolds.\\
 \includegraphics[width=84mm]{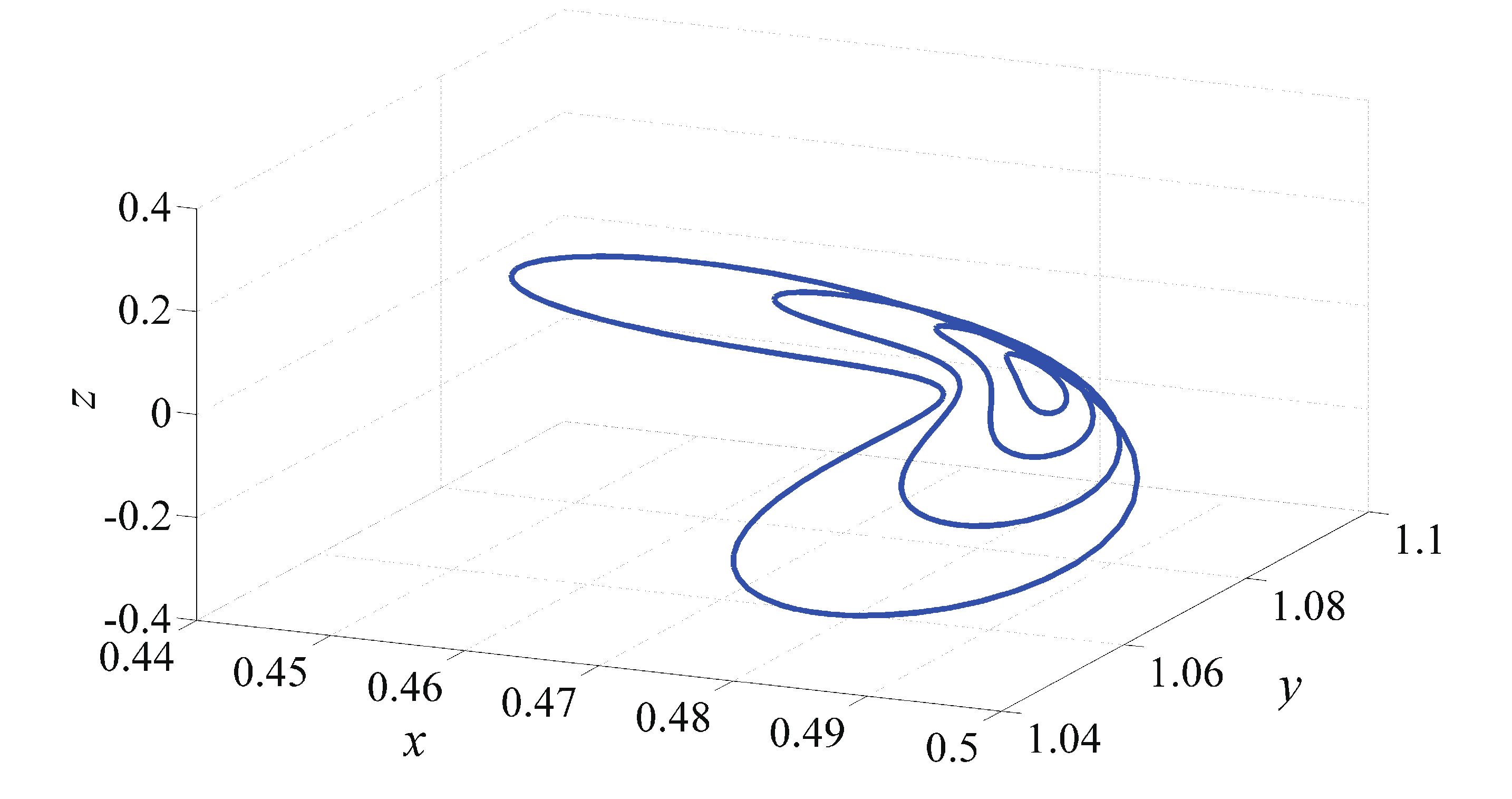}
 b. One 2-dimensional center manifolds of E2.\\
 \caption{Topological structure of manifolds of E2.}
\end{figure}
\begin{figure}
 \includegraphics[width=84mm]{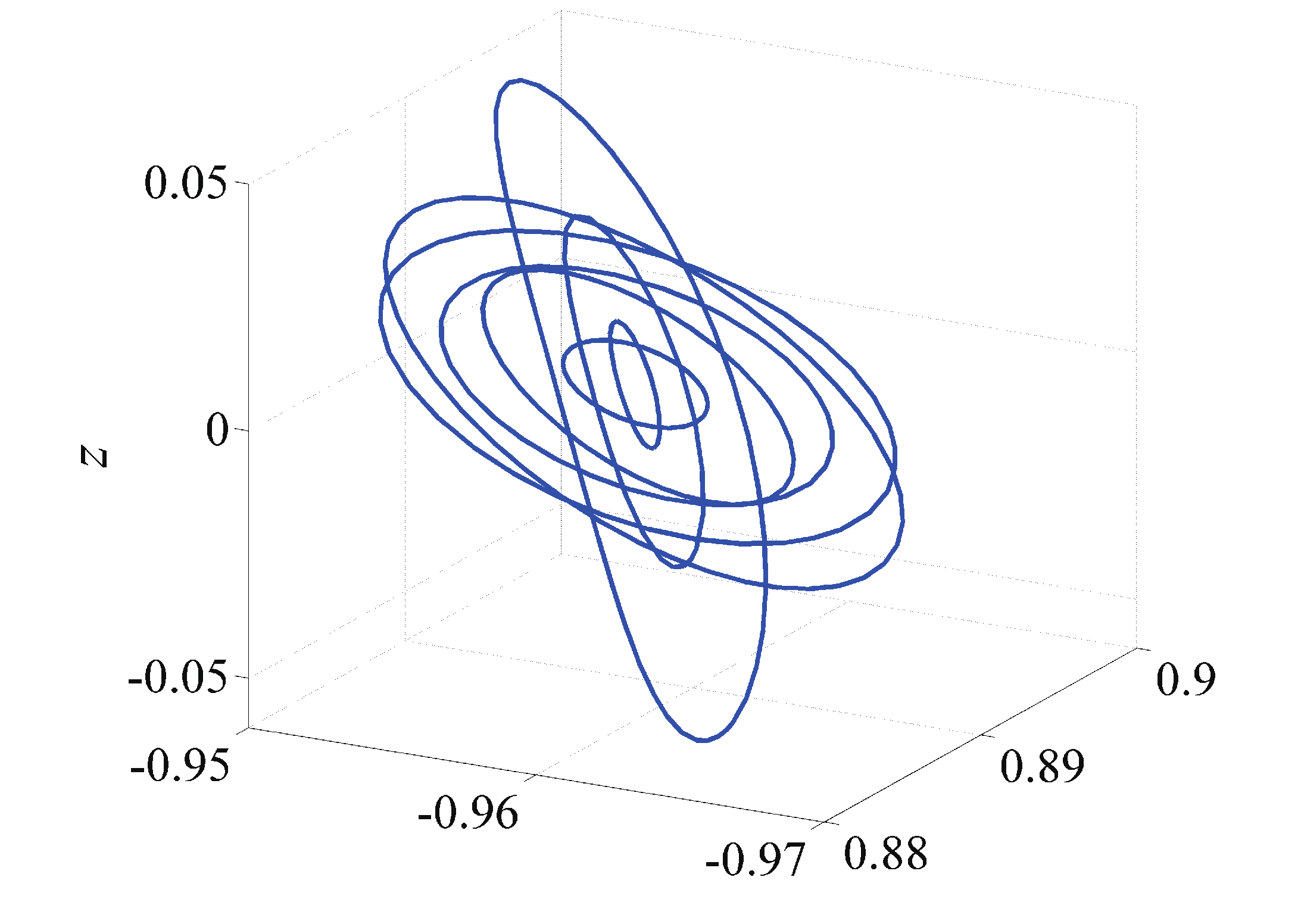}
 \caption{Topological structure of manifolds of E8: Three 2-dimensional center manifolds of E2. }
\end{figure}

Previous research has found that there are eight equilibrium points outside the rotating homogeneous cube, and this is very similar to the situation of Bennu, though the stability of the equilibrium points are different \citep{b10}.

\section{BIFURCATION OF THE EQUILIBRIUM POINTS}

Many asteroids in the solar system have a shape similar to Bennu, such as 1999 KW4 and 1950 DA. The dynamical structure is similar among these small bodies. Using the shape model of Bennu and the nondimensionalized method, the results of the simulation presented in this paper can be translated to those asteroids that have the same ¡°spinning-top¡± shape. Apart from the polyhedral shape model, the dynamical equations in the vicinity of the irregular-shaped small bodies are only relevant to the physical properties of the small bodies, such as the density and rotation period. Equations (18), (20), and (22) show that these physical properties are all involved in the dimensionless quantity $ \eta$. Increasing the bulk density and decreasing the rotation velocity have the same effects on the dynamical equations. Asteroids with different sizes, rotation periods, and bulk densities may have the same dynamical structure near the body if they have the same dimensionless quantity $ \eta$. Therefore, the dimensionless quantity $ \eta$ can be changed to investigate the influence of the variation of the physical properties on the dynamical environment near an asteroid. In this study, the dimensionless quantity $ \eta$ is varied from 0.3 to 1.0, and the bifurcation of both the number of equilibrium points and the topological structure of the manifold near the equilibrium points are presented.

\subsection{Bifurcation of the number of equilibrium points}

Generally, the number of the equilibrium points is thought to be constant with the variation of the system parameters, such as the homogeneous cub case or the circular restrict three body problem. Previous researchers have also investigated the change of the dynamical system using variable physical properties. \citet{b8} investigated the failure modes of the asteroid 216 Kleopatra using a variable size and bulk density with a constant mass. The mass and shape are given by \citet{b5} and \citet{b13}, respectively. The scale size had been changed from 1.0 to 1.5 with Ostro's size as its unit value. Hirabayashi found that the asteroid body was structurally stable when the scale size is between 1.18 and 1.32. \citet{b19} used a Ridge line to investigate the equilibrium points and the manifold near them with variable bulk density. By introducing a $ z^*$ set and a $ zh^*$ set, the problem of finding the equilibrium points is reduced from three-dimensions to one-dimension. This method clearly gives the equilibrium points and simplifies the problem greatly.

The equilibrium points can be solved using the equations in cylindrical form
\begin{equation}
   \nabla \tilde{V} (h, \theta, z) =0,
\end{equation}
where $ h$, $ \theta $, and $ z $ represent the radial distance, azimuth, and height coordinates respectively. The $ zh^*$ set was then defined as
\begin{equation}
   zh^*=\left\{ \tilde{\bmath{r}} \in \bld{R}^3| \tilde{V}_h (\tilde{\bmath{r}}) = 0, \tilde{V}_z (\tilde{\bmath{r}}) = 0 \right\},
\end{equation}
which has been proven as a smooth manifold of dimension 1. Generally, the $ zh^*$ set consists in a closed smooth curve, called Ridge Line.

Solving Equation (36) gives the equilibrium points of the potential field. We change the dimensionless $ \eta$ from 0.3 to 1.0, and each time solve Equation (36) to obtain the equilibrium points. Because of the irregular shape of Bennu, there is a variation in the number of equilibrium points.

Fig. 8 shows the variation of the number of equilibrium points of the asteroid 101955 Bennu. The number and position of the equilibrium points of Bennu both change if the dimensionless quantity $ \eta $ is varied. As the dimensionless quantity $ \eta $ increases, the equilibrium points E7 and E8 on the bottom right of Fig. 5(a) mix and disappeared as shown in Fig. 8(d). The same situation occurs for the equilibrium points E3 and E4 on the upper left of Fig. 8(a) if $ \eta $ continues to increase, as shown in Fig. 8(f). Finally, there are only five equilibrium points, four outside the body and one inside. As $ \eta$ increases, only the position of the equilibrium points will change, while the number of equilibrium points remains unchanged.
\begin{figure*}
\subfigure[$ \eta=0.30$]{
\includegraphics[width=160.00 pt]{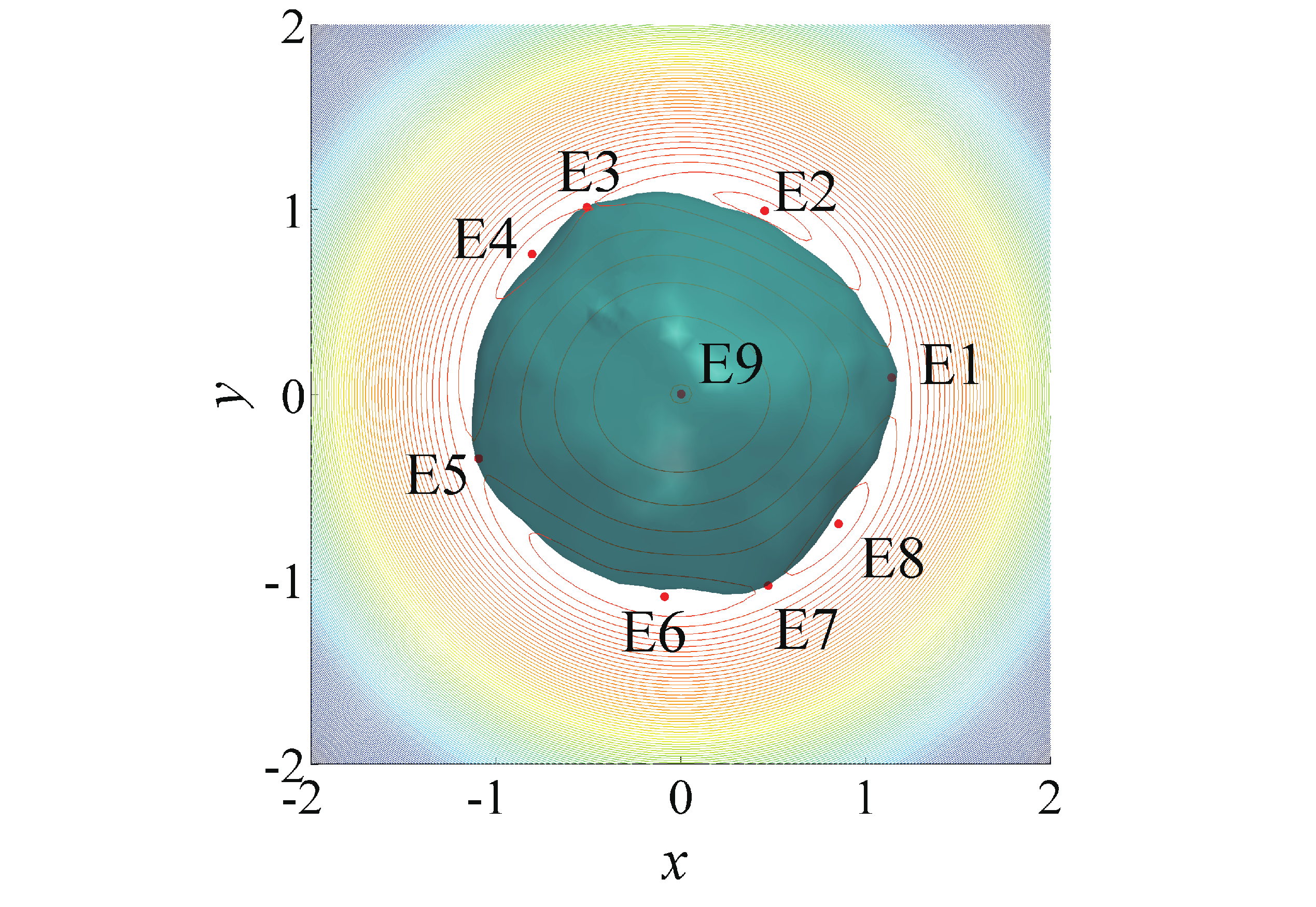}}
\subfigure[$ \eta=0.55$]{
\includegraphics[width=160.00 pt]{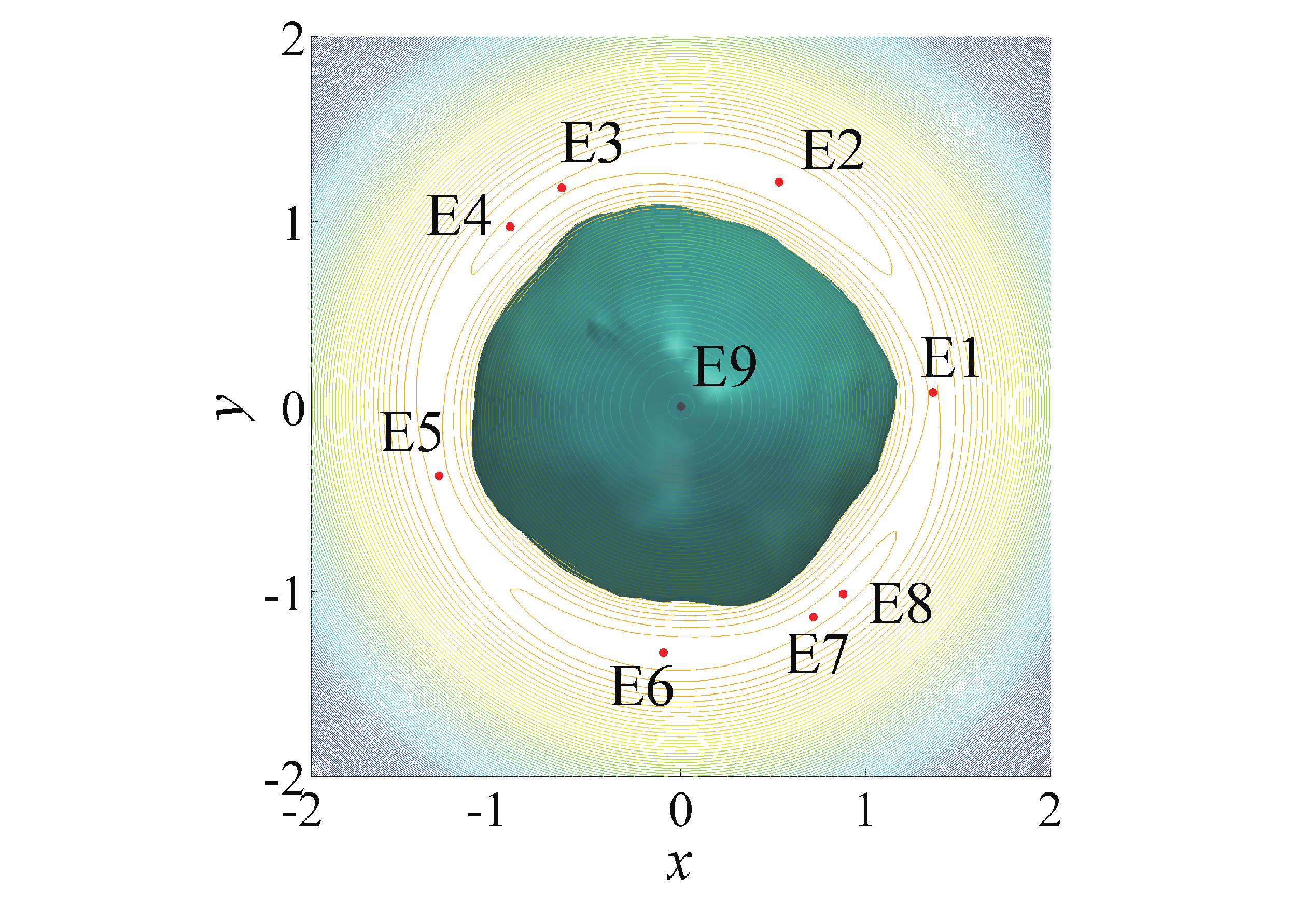}}
\subfigure[$ \eta=0.57$]{
\includegraphics[width=160.00 pt]{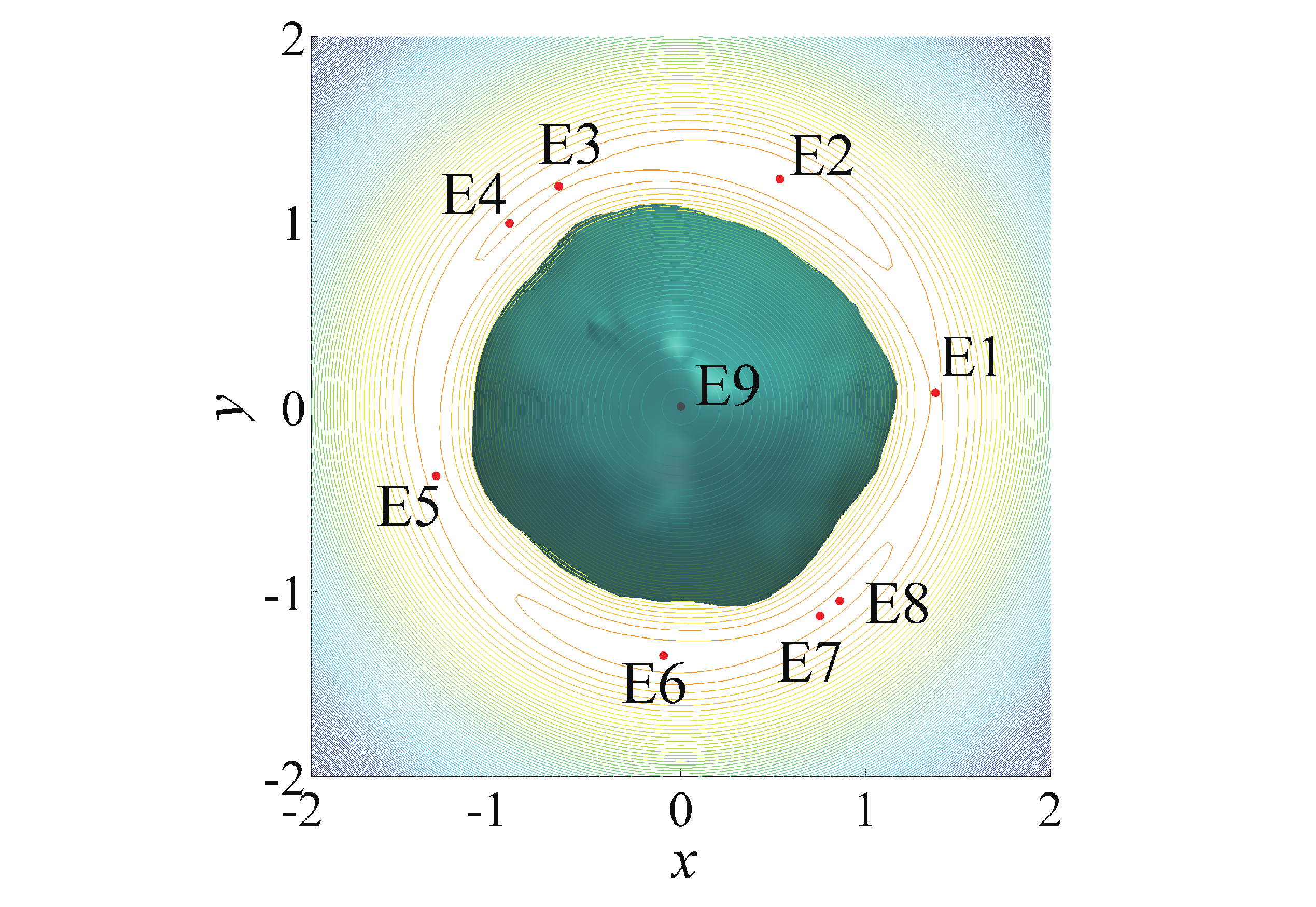}}
\subfigure[$ \eta=0.65$]{
\includegraphics[width=160.00 pt]{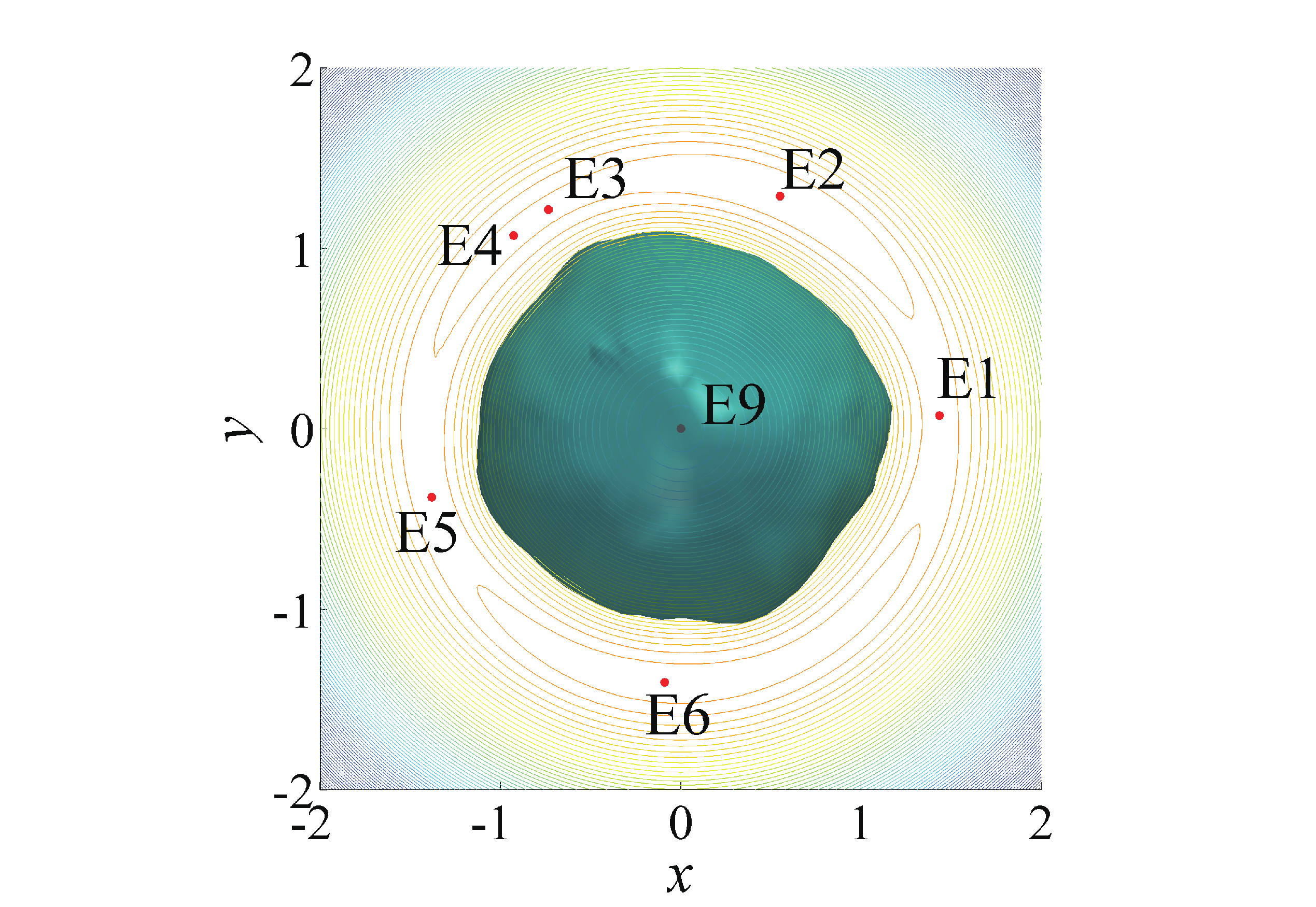}}
\subfigure[$ \eta=0.70$]{
\includegraphics[width=160.00 pt]{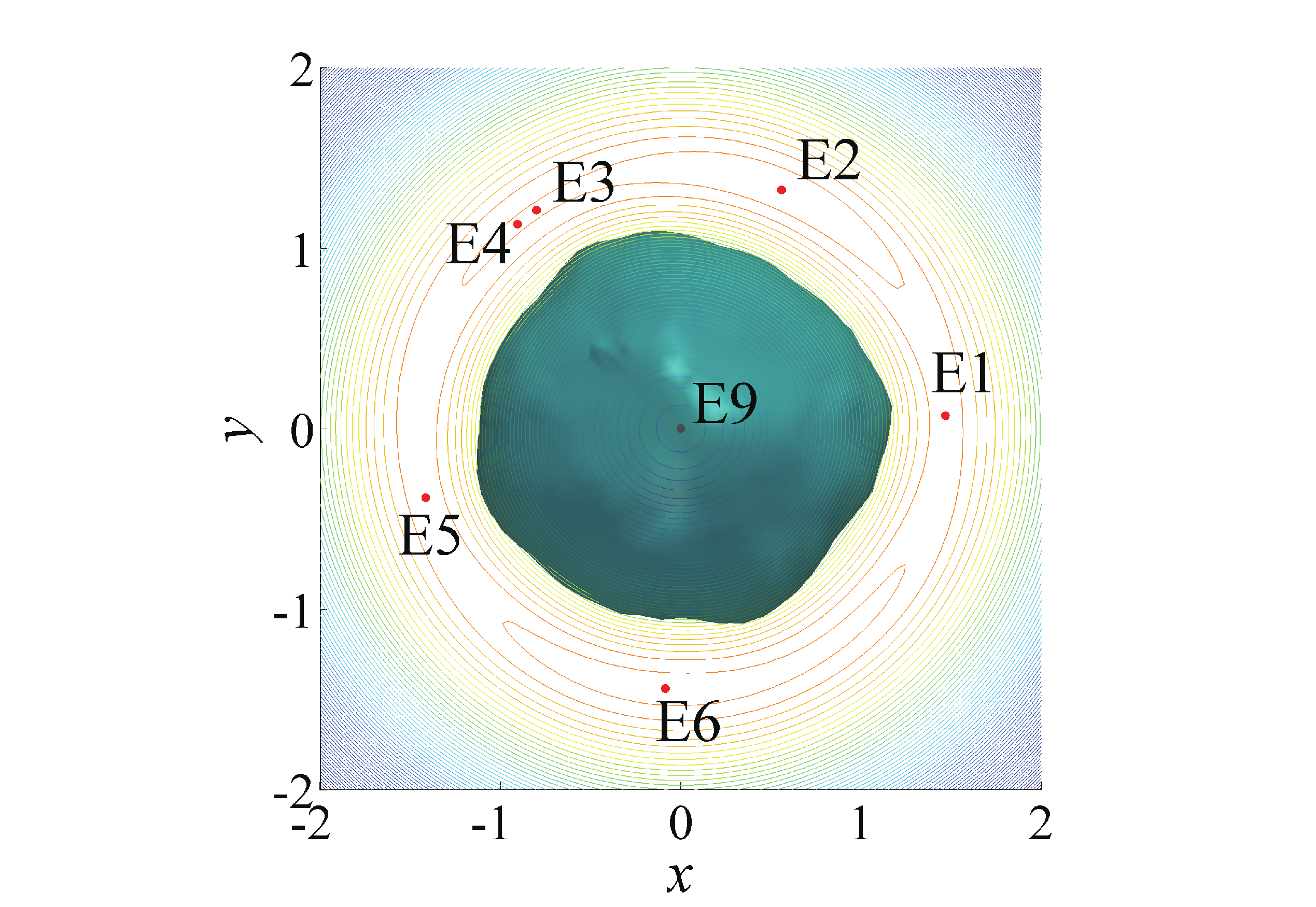}}
\subfigure[$ \eta=1.00$]{
\includegraphics[width=160.00 pt]{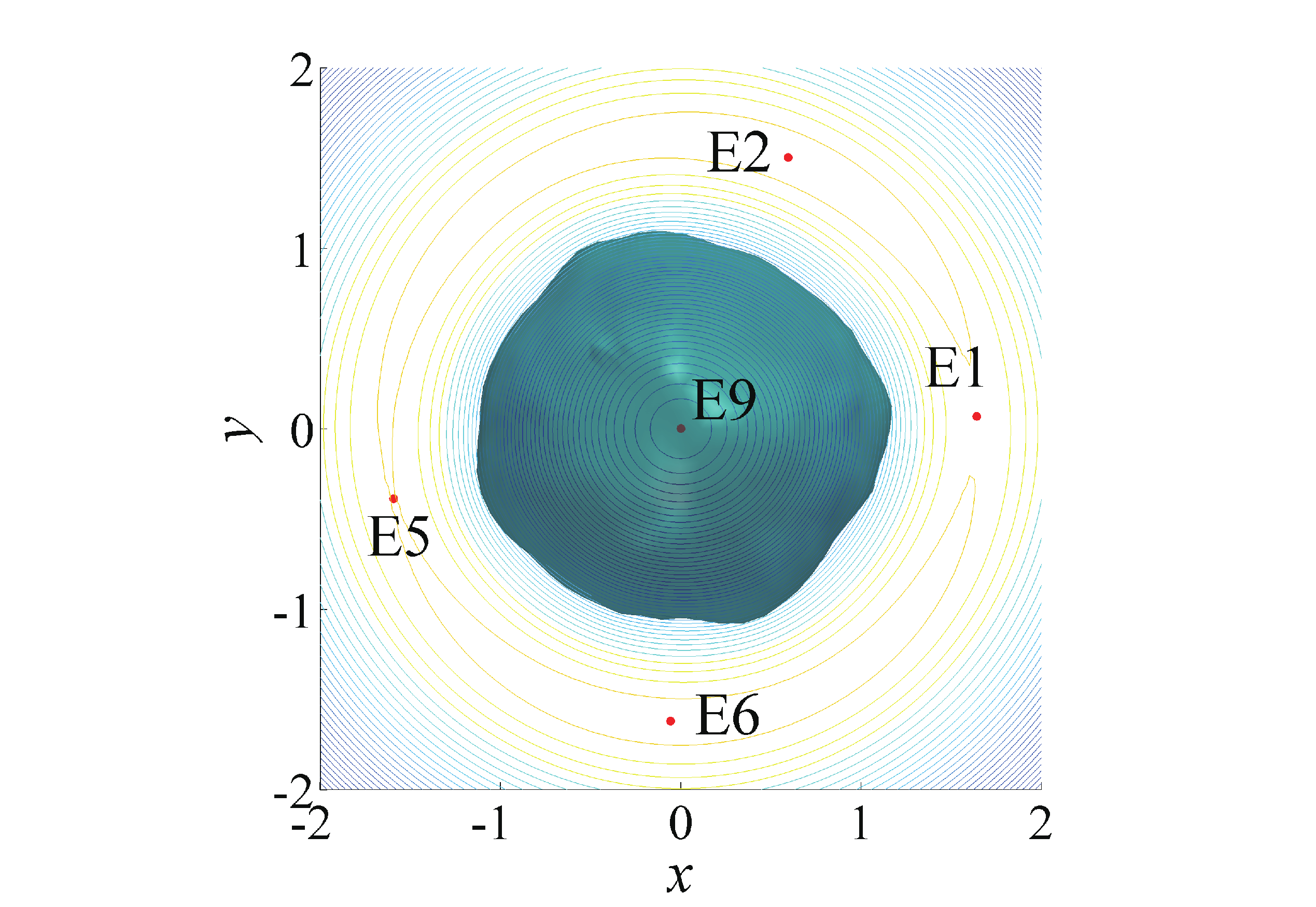}}
 \caption{The Equilibrium Points and Zero Velocity Curves of Bennu with the Variation of $ \eta$.}
\end{figure*}

\begin{figure*}
\subfigure[$ \eta=0.392 $]{
\includegraphics[width=160 pt]{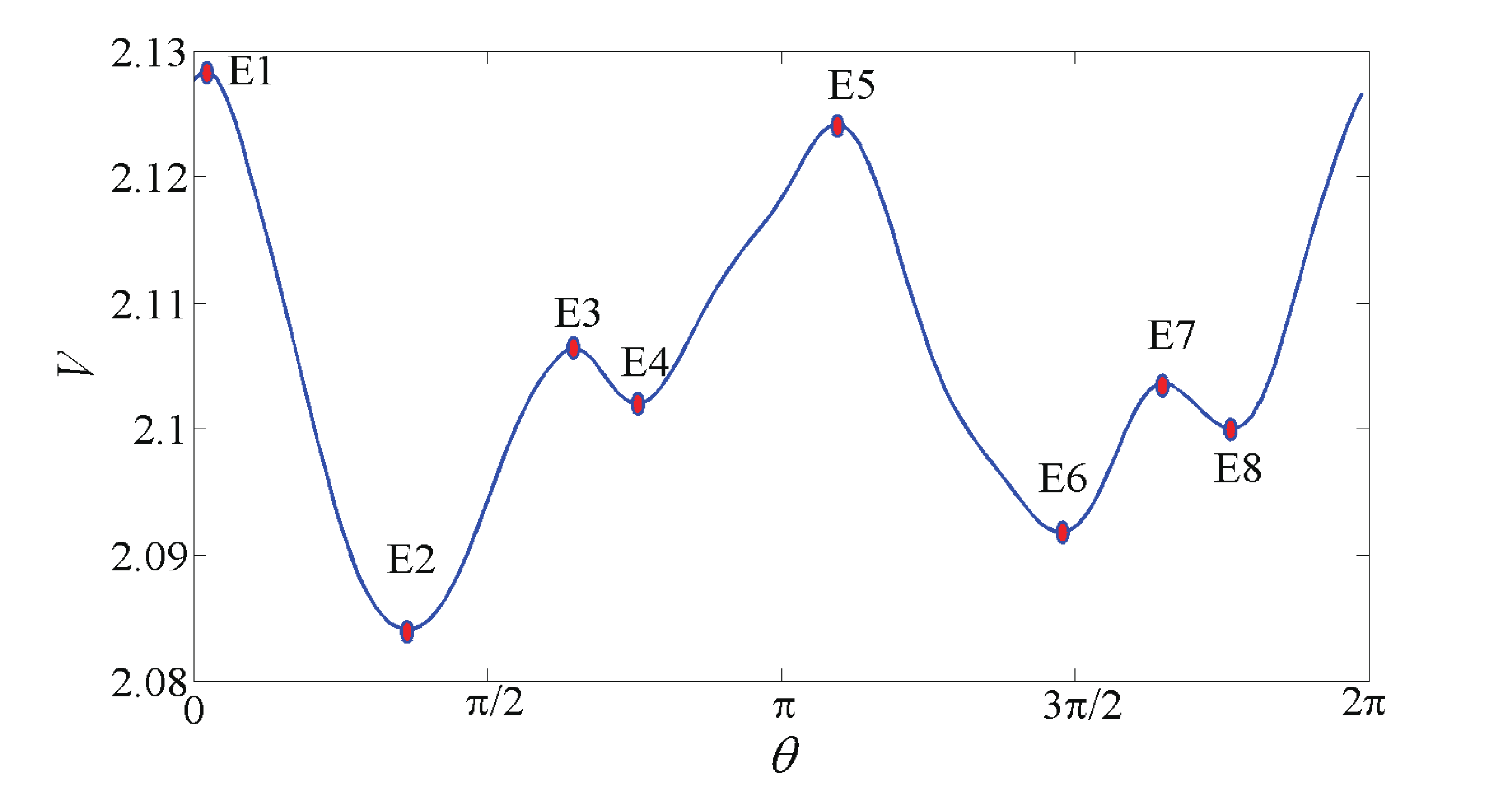}}
\subfigure[$ \eta=0.529 $]{
\includegraphics[width=160 pt]{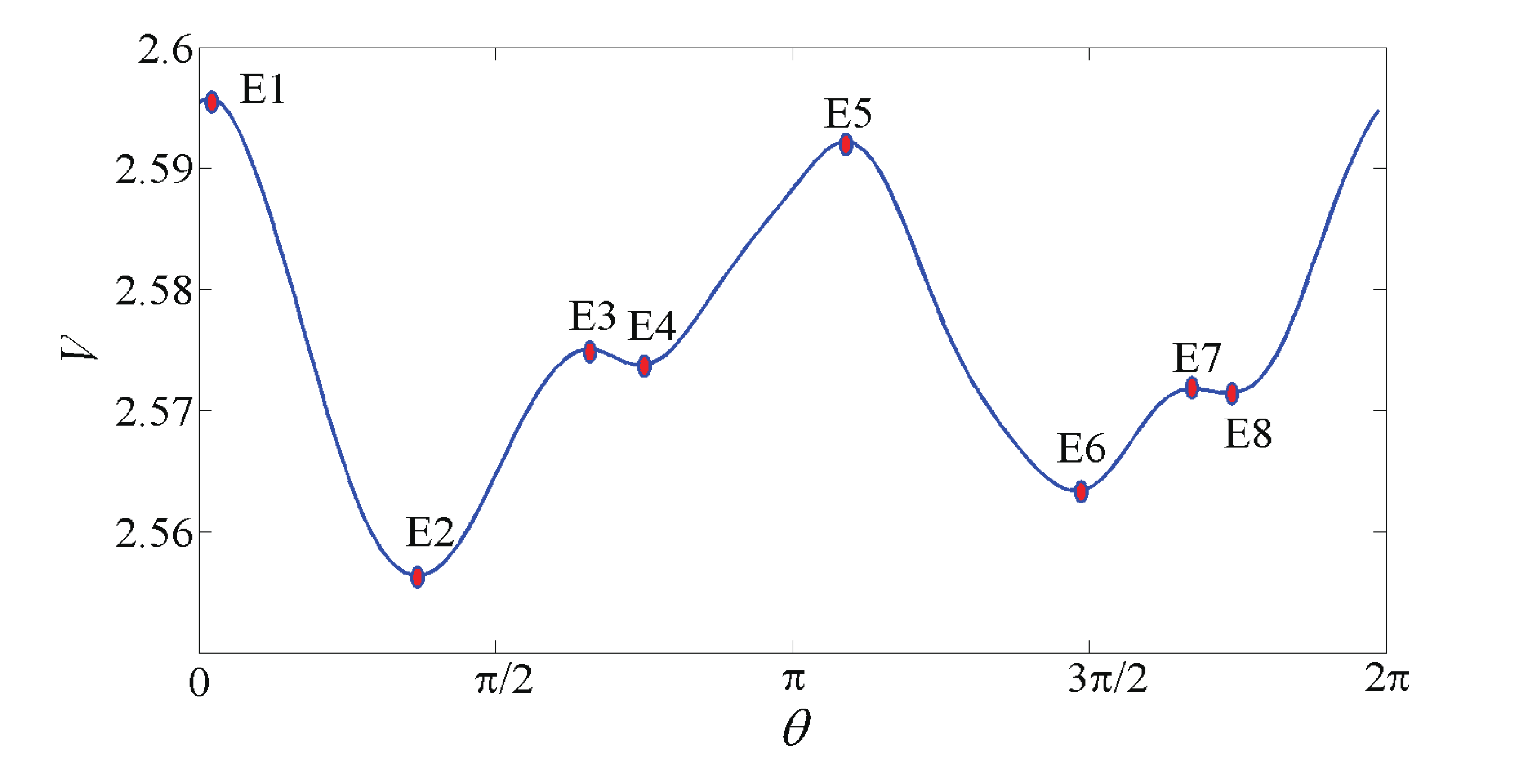}}
\subfigure[$ \eta=0.585 $]{
\includegraphics[width=160 pt]{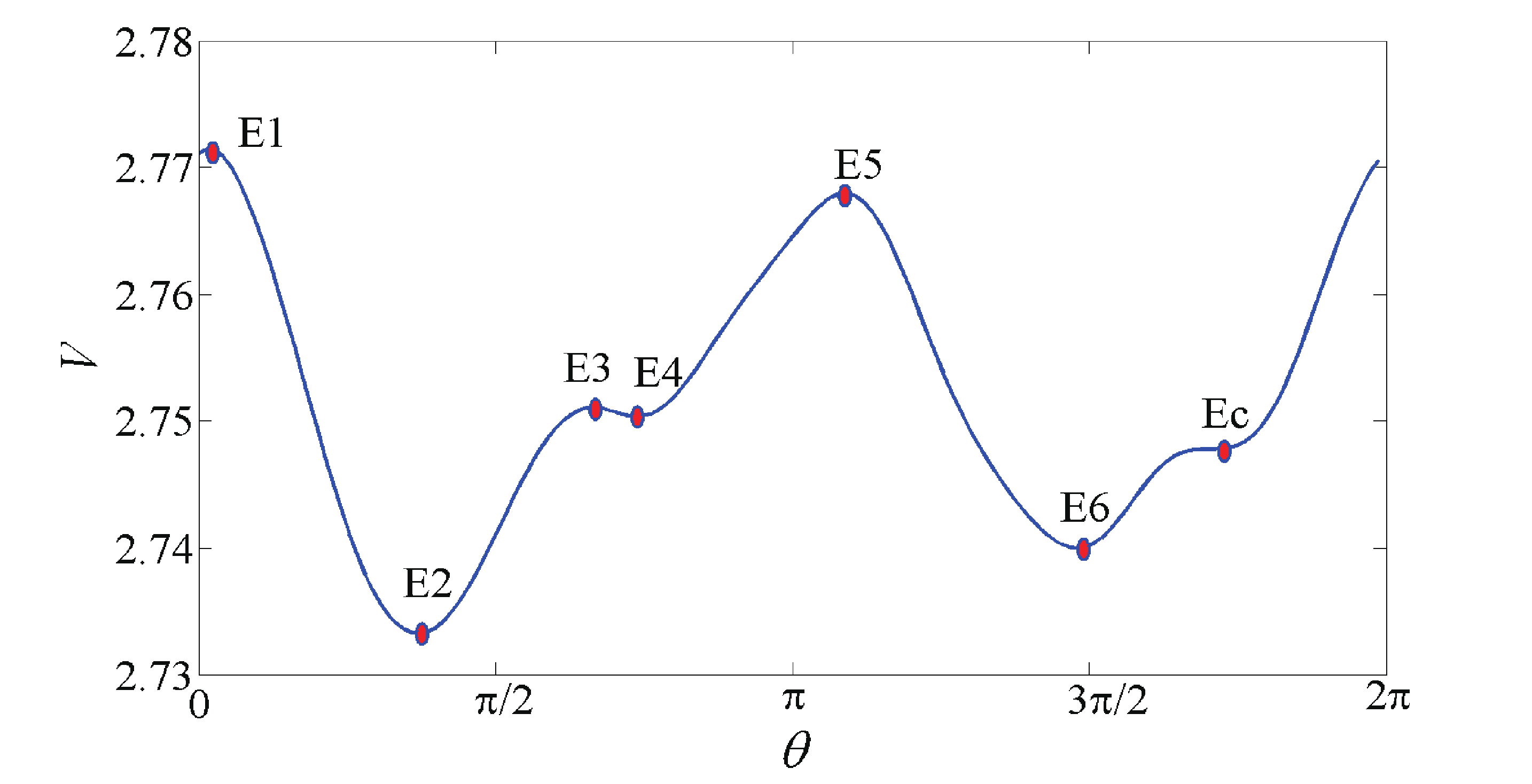}}
\subfigure[$ \eta=0.715 $]{
\includegraphics[width=160 pt]{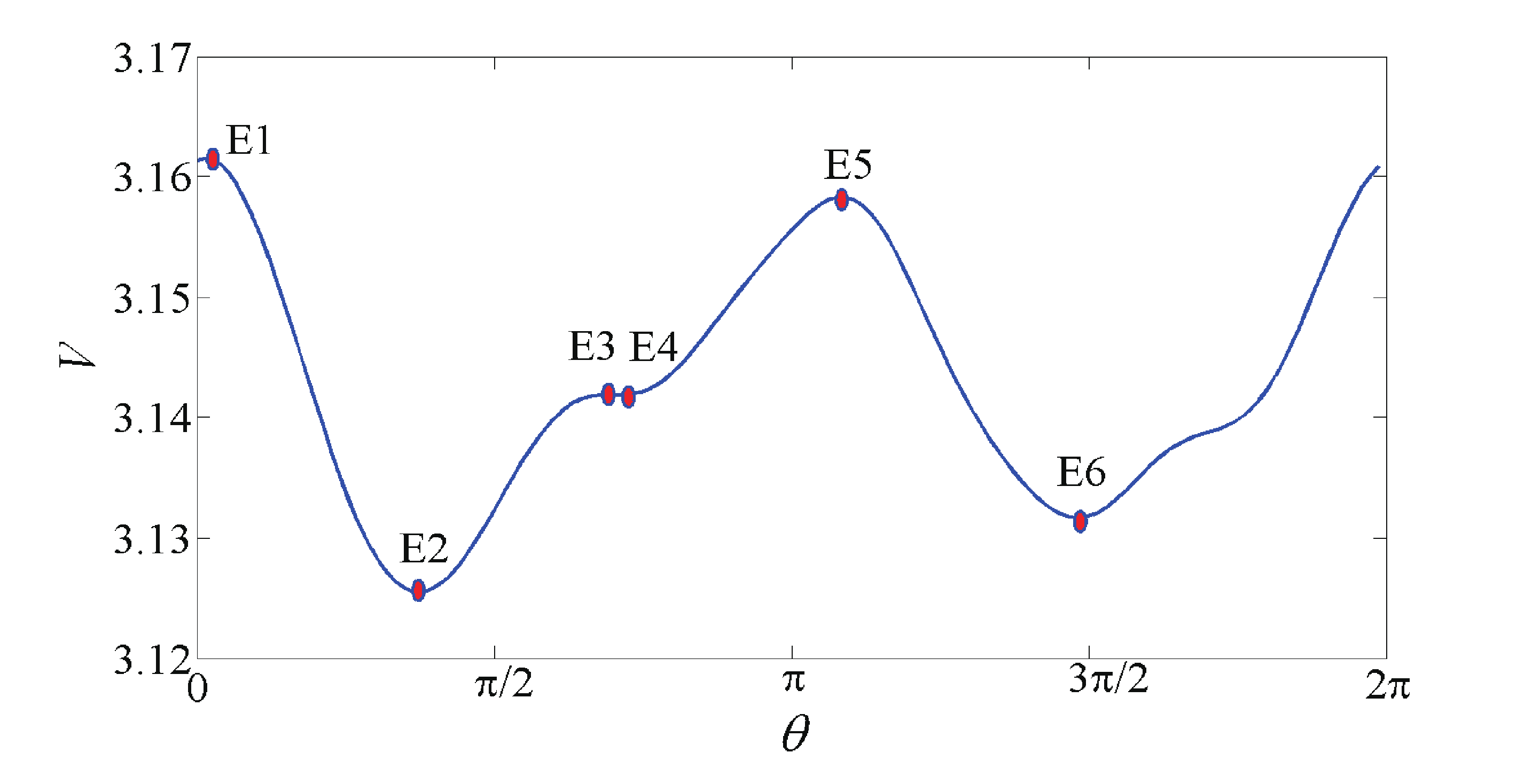}}
\subfigure[$ \eta=0.719 $]{
\includegraphics[width=160 pt]{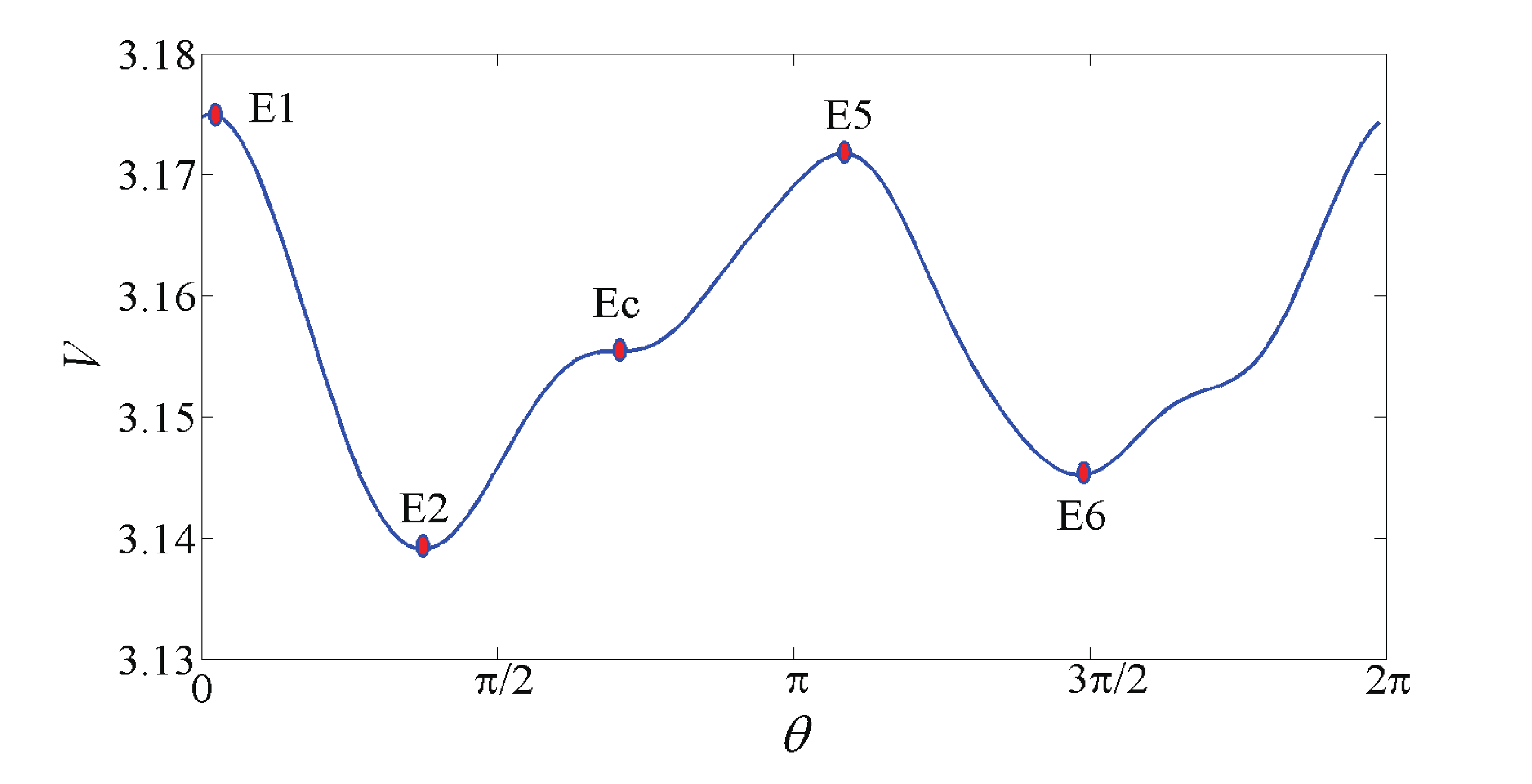}}
\subfigure[$ \eta=0.855 $]{
\includegraphics[width=160 pt]{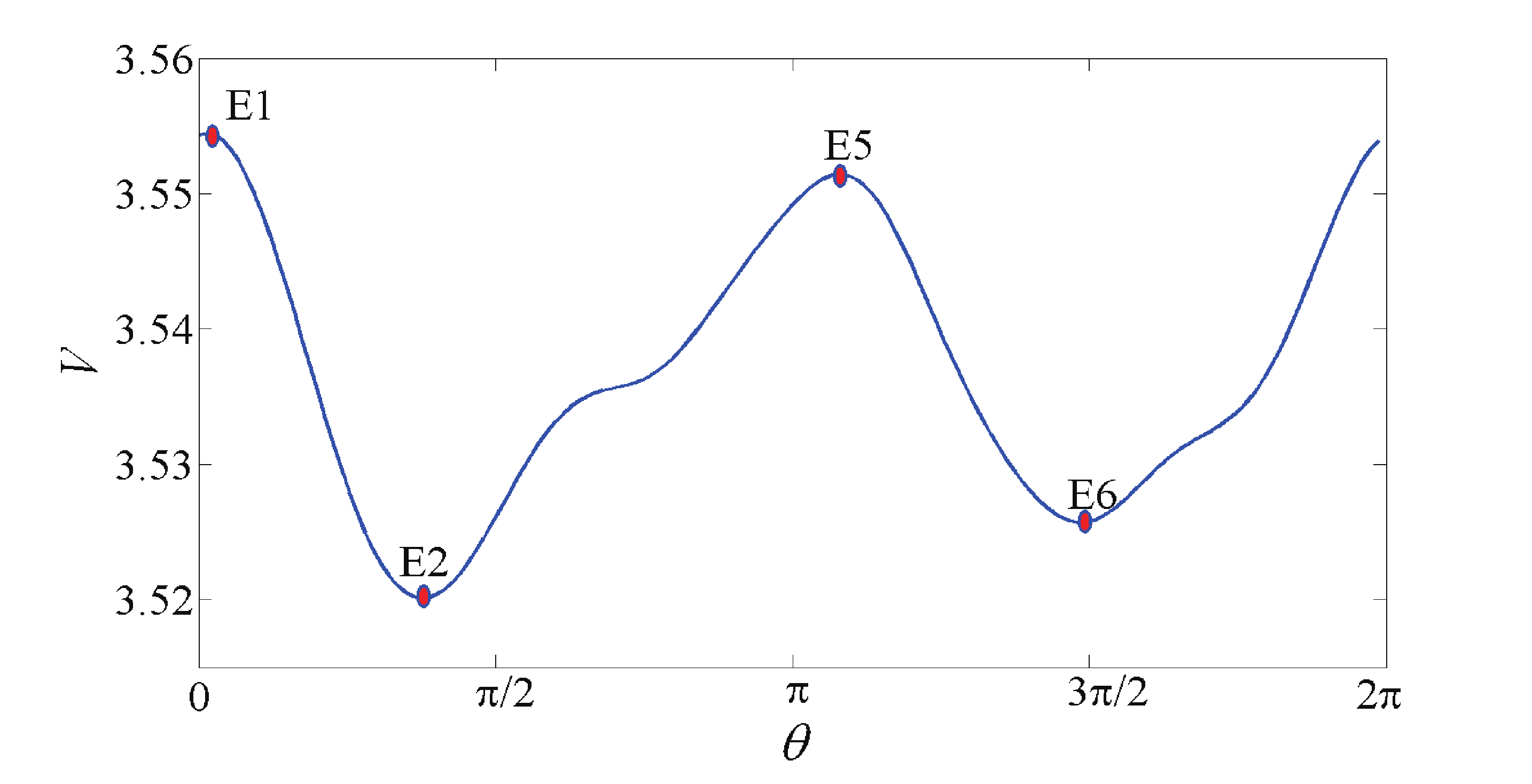}}
 \caption{Efficient potential $ V (\theta)$ in the Ridge Line with the different of $ \eta $.}
\end{figure*}

The results can be demonstrated more clearly using the $zh^*$ set method. We can calculate the ridge line with the variation of the dimensionless quantity $\eta$. In the Ridge Line, the efficient potential is only related to the azimuth $\theta$, i.e., $V =V (\theta)$. The equilibrium points can then be obtained using $V_{\theta}(\theta) = 0$. Fig. 9 shows the efficient potential $V (\theta)$ in the Ridge Line for different values of $\eta$. Generally, the peaks and valleys in Fig. 9 represent the equilibrium points that satisfy $V_{\theta} (\theta) = 0$. We can see that the peak and valley, which indicate E7 and E8, get smoother as $\eta$ increases. The peak has been sliced off, the valley has been filled up to the same level, and then they both disappear. Equilibrium points E7 and E8 get closer and merge into one equilibrium point. This result indicates that a critical situation only exists in a certain condition. The equilibrium point disappears with a small increase of the dimensionless quantity $\eta$. This critical situation can be easily proven as follows. Fig. 9(b) shows that the points in the Ridge Line between E7 and E8 have $V_{\theta} (\theta) < 0$ with $\eta=0.529$. These points have $V_{\theta} (\theta) > 0$ with $\eta=0.715$ in Fig. 9(d). Because the effective potential $V$ given by the polyhedral method is $C^1$ with $\eta$ and $\theta$ in the Ridge Line, i.e., $V(\theta, \eta) \in C^1$, there must be a critical situation that satisfies $V_{\theta}(\theta)=0$ among these points, which means that equilibrium points E7 and E8 touch each other. The same situation occurs between E3 and E4.

\begin{figure}
 \includegraphics[width=84mm]{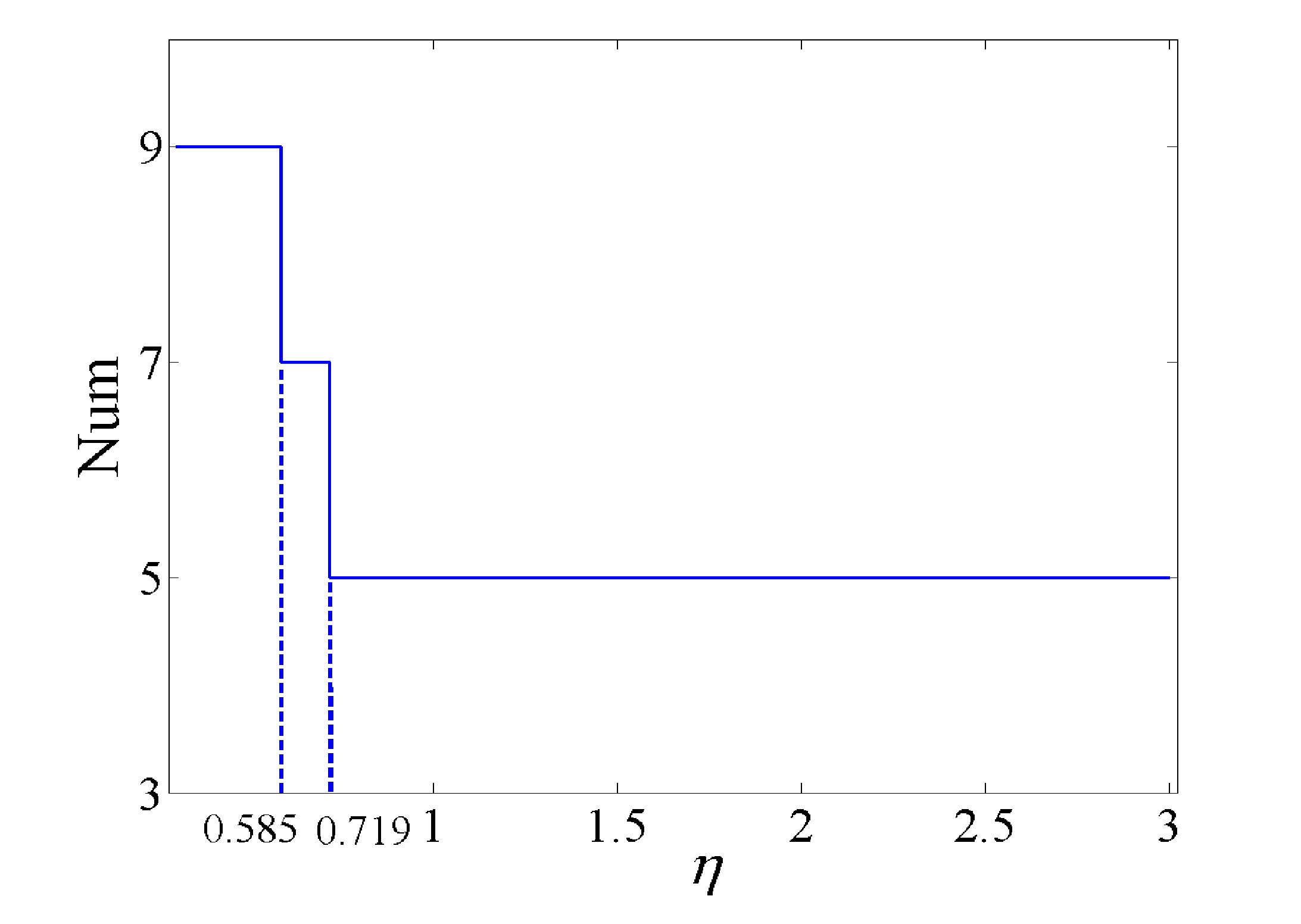}
 \caption{The Number of Equilibrium Points.}
\end{figure}

Fig. 10 shows the number of equilibrium points of Bennu with the variation of the dimensionless quantity $\eta $. The critical values of $\eta$ are $\eta_1 = 0.585$ and $\eta_2  = 0.719$ based on numerical computation. If $\eta$ is less than $\eta_1$, i.e., $\eta < \eta_1$, there will be nine equilibrium points in the potential field of Bennu, eight outside the body and one inside. If $\eta$ is less than $\eta_2$ and larger than $\eta_1$, i.e., $\eta_1 <  \eta < \eta_2$, there will be seven equilibrium points in the potential field of Bennu, six outside the body and one inside. Otherwise, if $\eta$ is larger than $\eta_2$, i.e., $\eta > \eta_2$, there will be five equilibrium points in the potential field of Bennu, four outside the body and one inside. If the dimensionless quantity $\eta$ is equal to its critical values $\eta_1$ or $\eta_2$, there may be eight or six equilibrium points in the potential field of asteroid Bennu. These are critical situations that hardly happen in reality.

\subsection{Bifurcation of the topological structure of the manifold near the equilibrium points}

The topological structure of the equilibrium points will also change, which indicates bifurcation. Corresponding to Fig. 8, the topological structure of the equilibrium points are shown in Table 4. The table shows that an equilibrium point can change from an unstable saddle point to a linear stable centre point, which indicates that there will be new period orbits near the equilibrium point.

\begin{table}
 \caption{The Topological Structure of the Equilibrium Points.}
 \label{symbols}
 \begin{tabular}{@{}lccccc}
  \hline
  \makecell{Equilibrium\\ Points}  & $\eta$=0.300 & $\eta$=0.500
        & $\eta$=0.526 & $\eta$=0.528 & $\eta$=0.553 \\
  \hline
  E1    &Case2 & Case2 & Case2 & Case2 & Case2\\
  E2    &Case5 & Case5 & Case5 & Case5 & Case1\\
  E3    &Case2 & Case2 & Case2 & Case2 & Case2\\
  E4    &Case5 & Case5 & Case1 & Case1 & Case1\\
  E5    &Case2 & Case2 & Case2 & Case2 & Case2\\
  E6    &Case5 & Case5 & Case5 & Case1 & Case1\\
  E7    &Case2 & Case2 & Case2 & Case2 & Case2\\
  E8    &Case5 & Case1 & Case1 & Case1 & Case1\\
  E9    &Case1 & Case1 & Case1 & Case1 & Case1\\
  \hline
 \end{tabular}
\end{table}

Fig. 11 shows the eigenvalue of the Jacobi matrix of equilibrium point E8 as an example. The dimensionless quantity $\eta$ is changed from 0.30 to 0.52 in this figure. The dot symbol in red indicates that the topological structure of equilibrium point E8 belongs to case 5, which has two pairs of conjugate complex numbers and one pair of pure imaginary numbers. The star symbol in blue indicates that the topological structure of equilibrium points E8 belongs to Case 1, which has three pairs of pure imaginary numbers. As $\eta$ increases, the two pairs of conjugate complex numbers get closer to the imaginary axis. In a critical situation, two upper complex numbers merge with each other to become one pure imaginary number, and the same happens for the lower complex numbers. Then, $\eta$ continues to increase, and the pure imaginary number splits into two pure imaginary numbers up and down the imaginary axis. Equilibrium points E2, E4, and E6 have a similar bifurcation progress, which is not shown here.

\begin{figure}
 \includegraphics[width=84mm]{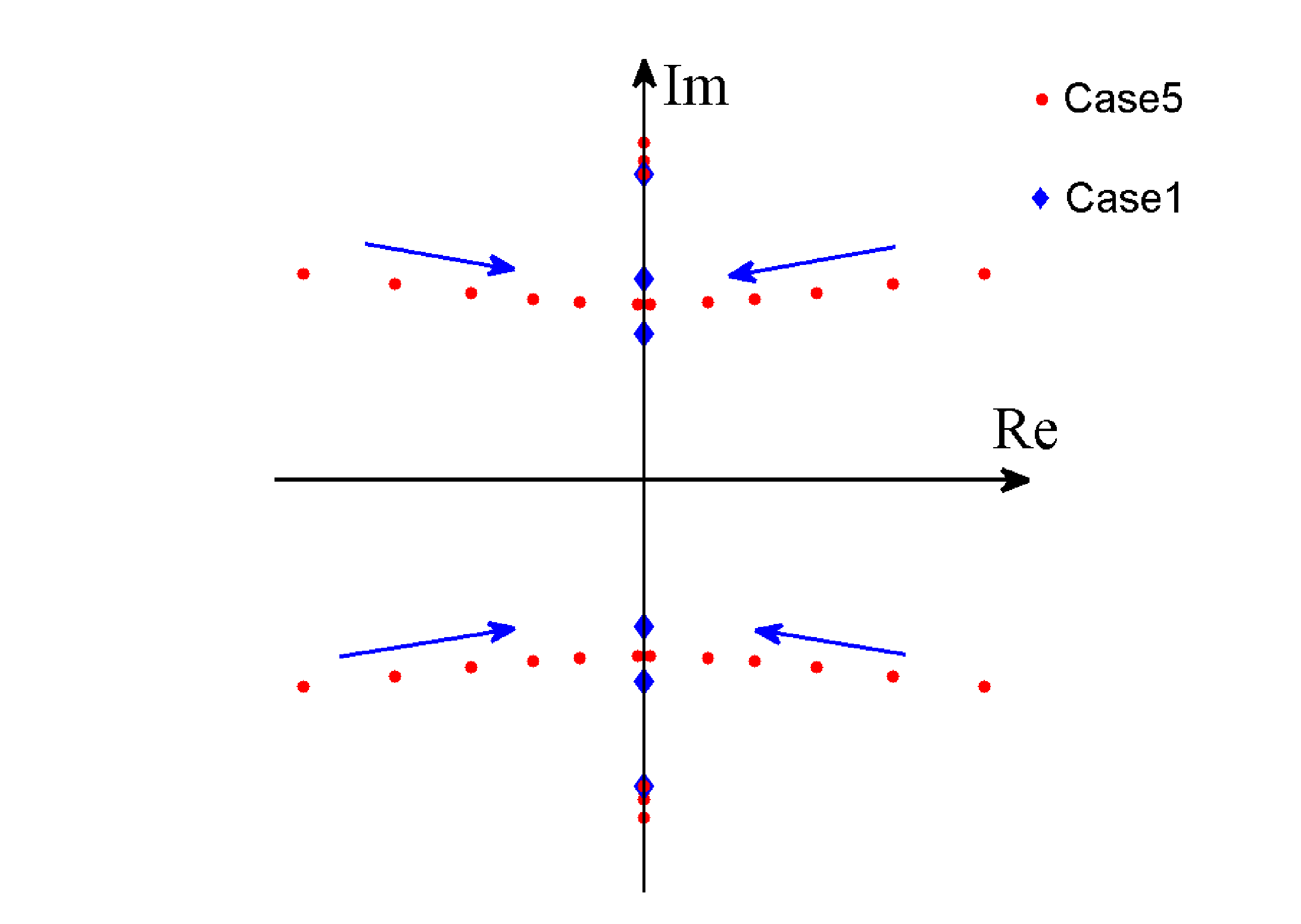}
 \caption{Eigenvalue of the Jacobi matrix of equilibrium point E8.}
\end{figure}

Using a bulk density of $1260  \pm  70 \rmn{kg m}^{-3}$ from \citet{b3} and a rotation period of $4.29746 \pm 0.002 \rmn{h}$ from \citet{b12}, the dimensionless quantity $\eta$ was calculated as $0.5095 \pm 0.0283$, i.e., $0.4812 \sim 0.5378$. Combining this with the analysis shown previously, we can obtain interesting results. Because the critical values of $\eta$ in the bifurcation of the topological structure are $\eta_3 = 0.500$, $\eta_4 = 0.526$, $\eta_5 = 0.528$, and $\eta_6 = 0.553$, when considering the error of the physical properties, the linear stability of the equilibrium points can be uncertain. E4, E6 and E8 can be either linear stable or unstable.

These results show the complexity of the potential field of Bennu. The variation of $\eta$ can change the stability of the equilibrium points. Additionally, the topological structure and period orbits also change. The uncertainty of the physical properties of Bennu means that the equilibrium points could be either unstable or linear stable, which is significant to know when designing a mission to orbit this small body.

\section{Conclusions}

In this paper, the stability and topological structure of the equilibrium points in the potential field of irregular-shaped minor bodies were investigated with variable density and rotation period, and the study shows fruitful results for the dynamical system of a noncentral rotating potential field. In order to gain a global view of the dynamical system of the minor bodies, both the internal and external potential fields were investigated.

Using the polyhedral method performed by \citet{b25}, we analysed the equilibrium points, their stability and period orbits. Moreover, the bifurcation of the number of equilibrium points and their topological structures are presented with the variation of the dimensionless quantity $\eta$. The results show that the equilibrium points are not immutable. The number of equilibrium points changes from nine to five as $\eta$ increases, and the topological structure of some of the equilibrium points changes from Case 5 to Case 1, which means that the equilibrium points changing from an unstable saddle point to a linear stable centre point. The critical values of the dimensionless quantity $\eta$, which indicate bifurcation, have also been found using a numerical method. The critical values relevant to the number of equilibrium points are $\eta_1=0.585$ and $\eta_2=0.719$. The critical values relevant to the topological structure of the equilibrium points are $\eta_3=0.500$, $\eta_4=0.526$, $\eta_5=0.528$ and $\eta_6=0.553$. Considering the measurement error of the bulk density from the ground-based observation \citep{b3}, the dimensionless quantity $ \eta$ of Bennu is $0.4812 \sim 0.5378$, which just covers some of the critical values. This result leads to uncertainty in the linear stability of the equilibrium points and indicates a complicated dynamical environment around Bennu. There can be many other conditions when applying this method to other irregular-shaped bodies yielding fruitful bifurcation.

This research provides an insight into the dynamic system in the vicinity of irregular-shaped small bodies using nonlinear dynamical theory, and the result can help us better understand the dynamical environment near small bodies such as asteroids and comets.

\section*{Acknowledgments}

This work is supported by the National Basic Research Program of China (973 Program) (2012CB720000).


\end{document}